\documentclass[12pt]{article}
\usepackage[tbtags]{amsmath}
\usepackage{amssymb}
\usepackage{graphicx}
\usepackage{color}
\usepackage{xcolor}
\usepackage[latin1]{inputenc}
\numberwithin{equation}{section}

\begin{document}
\title{Reaction Spreading in Systems\\ With Anomalous Diffusion}

\author{F. Cecconi\\
	CNR - Istituto dei Sistemi Complessi\\Via dei 
Taurini 19, I-00185 Rome, Italy \\ 
        D. Vergni\\
	CNR - Istituto Applicazioni del Calcolo\\
Via dei Taurini 19, I-00185 Rome, Italy \\
        A. Vulpiani\\
	Universit\`a di Roma ``Sapienza''\\ Piazzale A. Moro 5, 
I-00185 Rome, Italy.}

\vspace{0.5cm}
\maketitle

%=====================================================================
\abstract{ 
We briefly review some aspects of the anomalous diffusion, and its
relevance in reactive systems.  In particular we consider {\it strong
anomalous} diffusion characterized by the moment behaviour $\langle
x(t)^q \rangle \sim t^{q \nu(q)}$, where $\nu(q)$ is a non constant
function, and we discuss its consequences.  Even in the apparently
simple case $\nu(2)=1/2$, strong anomalous diffusion may correspond to
non trivial features, such as non Gaussian probability distribution
and peculiar scaling of large order moments.
 
When a reactive term is added to a normal diffusion process,
one has a propagating front with a constant velocity.
The presence of anomalous diffusion by itself does not guarantee 
a changing in the front propagation scenario; a key factor
to select linear in time or faster front propagation has been 
identified in the shape of the probability distribution tail
in absence of reaction.
In addition, we discuss the reaction spreading on graphs, underlying 
the major role of the connectivity properties of these structures, 
characterized by the {\em connectivity dimension}.  
}
%=====================================================================

%*******************************************************************
%DO NOT FORGET TO RESET THE EQUATION COUNTER TO 0 AT THE HEAD OF EACH SECTION
%*******************************************************************

%%%%%%%%%%%%%%%%%%%%%%%%%%%%%%%%%%%%%%%%%%%%%%%%%%%%%%%%%%%%%%%%%%%%%
\section{Introduction}
%%%%%%%%%%%%%%%%%%%%%%%%%%%%%%%%%%%%%%%%%%%%%%%%%%%%%%%%%%%%%%%%%%%%%
Transport processes and, in particular, diffusion play a very
important impact in science and technology \cite{Rosner_transp}, 
among the many examples
about their relevance we can mention their role in Geophysics,
e.g. pollutant spreading in atmosphere \cite{Atmo} or oceans \cite{Weiss}, 
in Biology and Chemistry as a framework to the interpretation of pattern 
formation and growth mechanisms \cite{Murray}, 
and finally in Mathematics for
their connection with stochastic processes \cite{Kampen}, including the  
historical importance of the Brownian motion for the
understanding of the physical reality of atoms
\cite{BM_Atoms}.

From a mathematical perspective, a transport process can be seen as a
class of deterministic or stochastic rule for the time evolution 
${\bf x} (0) \to {\bf x}(t)={\cal S}^t{\bf x}(0)$, where ${\bf x}$ belongs
to an unbounded domain and $|{\bf x}|$ typically increases
with time.  Likely, the paradigmatic
example of a transport process is given by a stochastic differential 
equation ruling
the time evolution of a fluid particle:
\begin{equation}
 \frac{d \mathbf x}{dt} = \mathbf u(\mathbf x,t)+
\sqrt{2D} \boldsymbol{\eta}(t) \;, 
\label{eq:langevin}
\end{equation}
where $\boldsymbol{\eta}(t)$ is a white noise, $D$ is the particle 
self-diffusivity and  $\mathbf u(\mathbf x,t)$ is a given velocity field 
driving the fluid, typically incompressible: $\nabla \cdot {\bf u}=0$.

We can associate to the above-mentioned dynamics an equation 
for the time evolution of the probability density function (PDF) 
$P(\mathbf x,t)$.
In this case one obtains the advection diffusion equation 
(also known as Fokker-Planck equation)
\begin{equation}
\frac{\partial P}{\partial t}+(\mathbf u \cdot \nabla) P = D \Delta P\, .
\label{eq:FP}
\end{equation}
Usually the explicit solution to Eq.~\eqref{eq:FP} cannot be worked out,  
however, for many purposes and applications, the knowledge of the 
asymptotic behaviour of the solution is sufficient to understand some 
basic properties of the transport process.
At large scale and asymptotically in time, the so 
called  {\it normal diffusion} is typically expected~\cite{Majda}, 
i.e. a Fick's law give an accurate approximation for the evolution
of ${\tilde P}$, that is the spatial coarse-graining of $P$,
\begin{equation}
\frac{\partial {\tilde P}}{\partial t}=
\sum_{i,j} D^e_{ij} \frac{\partial^2 {\tilde P}}{\partial x_i \partial x_j}\,
\label{eq:Fick}
\end{equation}
where 
$$
\langle x_i(t)x_j(t) \rangle \simeq 2 D^e_{ij} t \, .
$$
and $D^e_{ij}$ is the effective (eddy) diffusion 
tensor~\cite{EddyDiff}, \cite{Majda} depending, often in a not 
trivial way, on both $D$ and $\mathbf u$. We considered just for simplicity 
the case without average drift, $\langle{\bf x}\rangle =0$.
Eq.~\eqref{eq:Fick} implies a Gaussian behaviour for ${\tilde P}$:
\begin{equation}
{\tilde P}({\bf x}, t) 
\sim \exp 
\left\{-\frac{1}{4t} \sum_{i,j} 
x_i \left[D^e\right]^{-1}_{ij} x_j \right\}\,.
\label{eq:pdfFick}
\end{equation}
This common scenario is the simplest generalization of the
pure-diffusion case, where $\mathbf u=0$ and $D^e_{ij}=D\delta_{ij}$,
and the presence of the velocity field $\mathbf u$ has only a
quantitative effect for the values of the $D^e_{ij}$, but the Gaussian
behaviour is still valid.

Two questions naturally arise: a) is the normal diffusion 
a meaningful approximation?
b) how could the asymptotic behaviour of a transport process 
result in a violation of the Fick's law~\eqref{eq:Fick}?

We will discuss some features of the so called anomalous diffusion
where we have a failure of the previous scenario: $\langle x^2(t)
\rangle \sim t^{2 \nu}$ with $\nu \ne 1/2$ and a $P(\mathbf x, t)$
that is not Gaussian.  We will see that {\it anomalous diffusion} is a
consequence of the violation of the hypothesis of the central limit
theorem (CLT). It is important to remark that there are cases in which
anomalous behaviour can be rigorously proved (as in the case of
L\'evy-stable distribution).  Moreover we also consider the case of
{\it strong} anomalous diffusion, when $\langle x(t)^q \rangle \sim
t^{q \nu(q)}$ and $\nu(q)$ is a non constant function.  In addition,
we will also discuss the diffusion on discrete structure ``{\it
spatially}'' non homogeneous like graphs.  In such systems, e.g. in
the case of a ``fractal'' structure of the graphs, one can have
anomalous diffusion.

In chemistry, biology and physics it is rather common that, in specific
medium or environments, the transport phenomena are accompanied to, or even 
driven by, transformations or competitions among chemical or biological 
species~\cite{garcia2009}.  
Thus Eq.\eqref{eq:FP} needs to be modified by introducing some extra term 
describing such reactions.  
In this context, we can mention the simplest reaction-diffusion equation 
introduced by Fisher \cite{Fisher} and Kolmogorov, Petrovkii and
Piskunov \cite{KPP}: a one-dimensional system with normal diffusion and a 
reactive term:
\begin{equation}
{\partial \theta \over \partial t}=D{\partial^2  \over \partial x^2} \theta 
+{1 \over \tau} f(\theta)\,, 
\label{eq:FKPP}
\end{equation}
where the quantity $\theta(x,t)$ represents the concentrations of a chemical 
or biological species. 
The term $f(\theta) = \theta(1-\theta)$ describes an autocatalytic
reaction which allows for the transformation from the unstable 
state ($\theta = 0$) to the stable state ($\theta = 1$). 
In an unbounded domain, Eq.~\eqref{eq:FKPP} is known to admit 
travelling wave solutions moving at a constant speed 
$v_f=2 \sqrt{D f^{'}(0)/\tau}$~\cite{Fisher},~\cite{KPP}. 
 
One can wonder how the propagation behaviour of the system can change
when considering more general transport processes with a non-normal
diffusive behaviour, e.g., when either the Laplacian in
Eq.~\eqref{eq:FKPP} is replaced by another linear operator or the
dynamics spans the {\it ``spatially''} discrete structure of a
graph. We will see that the presence of anomalous diffusion is not
sufficient to induce a nonlinear front propagation in the presence of a
reaction.  Such a result can be understood, at least in some cases,
with a linear analysis starting from the shape of the $P(x,t)$ tail in
the reaction free system.

The paper is organized as follows: 
in Sect.~\ref{sec:diffusion} we briefly survey possible mechanisms 
that, violating the CLT hypothesis, 
lead to anomalous transport. In particular, we focus on 
processes where the anomalous behaviour is induced 
by: the statistical properties of the advecting fields,
the stochastic nature of the dynamics, or 
by restraining the random walks on specific graph-like structures. 
In Sect.~\ref{sec:front_Adiff}, we discuss the effect of the anomalous 
diffusive motion of reactants on the propagation front in reaction diffusion 
systems.  
In Sect.~\ref{sec:reacgraph}, we consider some cases where the 
anomalous transport is due to the graph-like geometry of the domain  
where the reaction diffusion takes place. 
Sect. 5 contains some conclusions.

%%%%%%%%%%%%%%%%%%%%%%%%%%%%%%%%%%%%%%%%%%%%%%%%%%%%%%%%%%%%%%%%%
\section{Mechanisms for anomalous diffusion} 
\label{sec:diffusion}
%%%%%%%%%%%%%%%%%%%%%%%%%%%%%%%%%%%%%%%%%%%%%%%%%%%%%%%%%%%%%%%%%

This section is devoted to the discussion of diffusion processes 
described by the superposition of many elementary displacements
\begin{equation}
\mathbf x(t+h) = \mathbf x(t) + h \mathbf v(\mathbf x(t),t). 
\label{eq:elem}
\end{equation}
where $h$ is a constant of the dimension of time, 
the term $\mathbf v(\mathbf x(t),t)$, defining
the character of the displacement, embodies
both deterministic and stochastic elements, as 
for the case of Brownian particles advected by a velocity field.
Eq.~\eqref{eq:elem} includes also the case of the motion over complex discrete 
structures when the displacements are identified with the links of a graph.

We shall see through the following examples, how the statistical
properties of the dynamics generated by the rule~\eqref{eq:elem}
depends on the correlation of the elementary displacements
\cite{Gabrielli}.  When such correlations allow the central limit
theorem to be applicable, the dynamics obeys the normal diffusion,
when not, anomalous behaviours could take place. However as reported
in~\cite{Kong1989}, there exist cases in which linear mean square
displacement does not necessarily imply a Gaussian behaviour of the 
probability distribution.  In particular, we
consider hereafter cases in which this correlation is already
intrinsic in $\mathbf v(\mathbf x(t),t)$ and cases in which it is
imposed by the geometric structure of the graph which constitutes the
support to the displacements.

In the limit $h\to 0$, the elementary displacements become infinitesimal, 
thus, the evolution \eqref{eq:elem} turns out to be continuous in time and 
can be generally written as the stochastic differential 
equation~(\ref{eq:langevin}).

Equivalently, we can consider the evolution of the PDF of $\mathbf x$ 
\begin{equation}
   \partial_t P = 
   \hat{L} P\,,
   \label{eq:passive}
\end{equation}
where the linear operator $\hat{L}$ accounts for the global
transport properties of the system, i.e., the macroscopic
operator describing the composition of the microscopic
term ${\mathbf v}({\mathbf x}(t),t)$.

In Section~\ref{sec:2.1}, we firstly investigate the conditions under
which the transport process~\eqref{eq:elem} exhibits a normal
diffusion behaviour for large times, then we discuss different models
that can be used to describe transport processes focusing on examples
able to show anomalous diffusion.  In particular we start from
advection-diffusion models, where $\hat{L}=-\mathbf{u}
\cdot
\mathbf{\nabla}+D \Delta$, passing through transport in 2D Hamiltonian
system, and finally ending to the mesoscale effective diffusion 
models suitable to study transport properties in turbulent flow 
and heterogeneous media~\cite{Procaccia1985}, where 
$
\hat{L}=\frac{1}{r^{d-1}}
\dfrac{\partial}{\partial r}\left(k(r)r^{d-1}
\dfrac{\partial}{\partial r}\right)\,.
$
In Section~\ref{sec:2.2} stochastic processes, and in particular
continuous time random walk are investigated as an example of transport 
dynamics showing strong anomalous diffusion characterized by the behaviour  
$\langle |x(t)|^q \rangle \sim t^{q\nu(q)}$ of its moments, where $\nu(q)$
is a non constant function.
Then, in Section~\ref{sec:diffgraph}, diffusion processes in discrete
systems with a non homogeneous spatial structure are taken into
account.
%Other examples of transport process~\eqref{eq:elem} in continuous
%time is
%or in discrete time,
%$$
%\mathbf x(n+1) = \mathbf f(\mathbf x(n))+
%\mathbf w(n)\,,
%$$ 
%where $\mathbf w(t)$ and $\mathbf z(t)$ are independent and
%identically distributed (i.i.d) Gaussian processes that are delta
%correlated in time. 

%>>>>>>>>>>>>>>>>>>>>>>>>>>>>>>>>>>>>>>>>>>>>>>>>>>>>>>>>>>>>>>>>>
\subsection{Anomalous diffusion in presence of velocity fields}
\label{sec:2.1}
%>>>>>>>>>>>>>>>>>>>>>>>>>>>>>>>>>>>>>>>>>>>>>>>>>>>>>>>>>>>>>>>>>
Here we consider the evolution of a test Brownian particle advected by
the incompressible velocity field $\mathbf u$, described by the
stochastic differential equation~(\ref{eq:langevin}).  Let us recall
the Taylor-Kubo relation~\cite{Majda} between diffusion coefficient
and velocity-velocity correlation function: a direct integration of
equation~(\ref{eq:langevin}) provides easily the mean square
displacement:
\begin{equation}
\langle \left[x_i(t)-x_i(0)\right]^2 \rangle = \int_0^t dt_1 \int_0^t dt_2 
\langle v_i(\mathbf x(t_1)) \; v_i(\mathbf x(t_2))\rangle 
\simeq 2\; t\; \int_0^t 
d\tau \; C_{ii} (\tau)\,,
\label{21}
\end{equation}
where 
\begin{equation}
C_{ij}(\tau)=\langle \; v_i ({\mathbf x}(\tau))\; v_j ({\mathbf x}(0)) \; \rangle
\label{22}
\end{equation}
is the correlation function of the Lagrangian velocity in 
Eq.\eqref{eq:langevin}) and ${\mathbf v} = \dot{\mathbf x}$
(see Eq.~\eqref{eq:elem}).
If the hypothesis of the central limit theorem are satisfied:
\begin{itemize}
   \item[a)] finite variance of the velocity, i.e., 
             $\langle|\mathbf v|^2 \rangle < \infty$,
   \item[b)] fast enough decay of 
$R_{ii}(\tau)=C_{ii}(\tau)/\langle{v_i^2}\rangle$, 
             i.e, $0<\int_0^t \rm d\tau \; R_{ii}(\tau) < \infty$,
\end{itemize}
a normal diffusion takes place and we can define an 
eddy diffusivity as
\begin{equation}
   D_{ii}^e = \lim_{t \to \infty} {1 \over 2t}    
         \langle{|x_i(t)-x_i(0)|^2}\rangle = 
     \langle{v_i^2}\rangle\int_0^\infty {\mathrm d} \tau \; R_{ii}(\tau)\,.
   \label{eq:d11}
\end{equation}
The previous considerations are still valid for test particle
evolving according to a discrete time map 
$${\mathbf x}(n+1) = {\mathbf f}({\mathbf x}(n))\,;$$
with the obvious changes 
\begin{equation}
\mathbf v \Rightarrow {\mathbf f}({\mathbf x}) - {\mathbf x} 
\qquad\mbox{and}\qquad
\int_0^t {\mathrm d} \tau C_{ii}(\tau) \Rightarrow {C_{ii}(0)\over 2} +
\sum_{k=1}^n C_{ii}(k)\,\,.
\label{andiscrete}
\end{equation}
On the contrary, if at least one of the hypothesis of the CLT fails,
the transport process could not be a normal diffusion. 
Therefore a finite $D_{ii}^e$ cannot be defined: 
formally $D_{ii}^e~=~\infty$ (superdiffusion) 
or $D_{ii}^e~=~0$ (subdiffusion). The reader should note that
super diffusion appears when either the integral $\int_0^t 
d\tau \; C_{ii} (\tau)$ in Eq.~\eqref{21} 
or the sum $\sum_{k=1}^n C_{ii}(k)$ in Eq.~\eqref{andiscrete}, 
diverges.

A very important result about anomalous diffusion in
incompressible velocity fields was obtained in~\cite{Avellaneda1991}
and \cite{Avellaneda1995} and can be summarized as follows. 
If the molecular diffusivity $D$ is non zero, and the infrared contribution 
from the Fourier components $\hat{\mathbf u}(\mathbf k)$ of the velocity field 
are weak enough, namely
\begin{equation}
\int d\mathbf k \; \frac{\langle|\hat{\mathbf u}(\mathbf k)|^2\rangle}
{k^2} < \infty\,,
\label{210}
\end{equation}
then the advected fluid particles undergo normal diffusion, i.e., 
the effective diffusion coefficients $D_{ii}^e$'s in (\ref{eq:d11}) are finite. 
The symbol $\langle \; \cdot \; \rangle$ indicates time average.  
There are two possible causes for the superdiffusion:\\ 
\indent a) $D > 0$ and, in order to violate the condition \eqref{210}, 
$\mathbf u$ with strong spatial correlation;\\ 
\indent b) $D=0$ and strong time correlation between 
$\mathbf u(\mathbf x(t))$ and $\mathbf u(\mathbf x(t+\tau))$ at large $\tau$.\\
In the following sub-sections we present three different
paradigmatic models showing anomalous diffusion originated
by different mechanisms:\\
i) transport in incompressible random flows (Sect.~\ref{sec:2.1.1}) where the 
failure of the standard diffusion relies on the breakdown of hypothesis 
\eqref{210};\\
ii) diffusion in symplectic maps (Sect.~\ref{sec:2.1.2}) that represents
a discrete version of transport process in 2D incompressible flows. 
In these system, the anomalous
behaviour is a consequence of the strong correlations at large times enhanced 
by the presence of ballistic-like trajectories;\\
iii) relative dispersion of two particles in three-dimensional turbulence
(Sect.~\ref{sec:2.1.3}), in which the Laplacian operator $\hat{L}$ 
is built by simple phenomenological arguments.

%>>>>>>>>>>>>>>>>>>>>>>>>>>>>>>>>>>>>>>>>>>>>>>>>>>>>>>>>>>>>>>>>>>>>>
\subsubsection{Random shear flow}
\label{sec:2.1.1}
%>>>>>>>>>>>>>>>>>>>>>>>>>>>>>>>>>>>>>>>>>>>>>>>>>>>>>>>>>>>>>>>>>>>>>
One of the few non trivial systems for which the presence of anomalous
diffusion can be proved rigorously is the $2D$ random shear
flow. In this case, the transport dynamics is generated by 
the combined effects of the velocity field and the molecular diffusivity. 
Matheron and De Marsily~\cite{Matheron} showed that
anomalous diffusion, in the $x-$direction, occurs in a 2D random shear
$
{\mathbf u}=(u(y),0)\;,
$ 
if the infrared contribution to $u(y)$ is strong enough to 
violate Eq.~(\ref{210}).
For example the infrared ($k \sim 0$) divergence,  
$\langle|\hat{\mathbf u}(\mathbf k)|^2\rangle \sim k^\gamma$
entails for $\gamma \geq 1$ a diffusive asymptotic 
transport; instead, for $-1 \leq \gamma < 1$, the transport is 
asymptotically superdiffusive: 
\begin{equation}
\langle{|x(t)-x(0)|^2}\rangle \simeq t^{2\nu}\qquad\mbox{with}\qquad
\nu = {3-\gamma \over 4} > 1/2\,.
\label{eq:diffanshear}
\end{equation}
The case $\gamma=0$ has been studied in great details~\cite{Bouchaud} 
and the PDF of a test particle around $x$ at time $t$ (starting with 
a distribution $P(x,0)$ localized around $x=0$) is
\begin{equation}
P(x,t) \sim \frac{1}{t^{3/4}} \exp 
\left[
-C\,{\left (|x| \over t^{\,3/4}\right)^4} 
\right]\,\,
\label{eq:pdfrsf}
\end{equation}
where $C$ is a constant.

%>>>>>>>>>>>>>>>>>>>>>>>>>>>>>>>>>>>>>>>>>>>>>>>>>>>>>>>>>>>>>>>>
\subsubsection{Anomalous diffusion in chaotic symplectic systems}
\label{sec:2.1.2}
%>>>>>>>>>>>>>>>>>>>>>>>>>>>>>>>>>>>>>>>>>>>>>>>>>>>>>>>>>>>>>>>>
Anomalous diffusion has been observed in 2D laminar periodically
time dependent incompressible flows~\cite{Castiglione}.
In such a case, the Poincar\'e map technique shows that 
the motion of a fluid particle is described by a 2D symplectic
map~\cite{Lichtenberg}. Therefore it is quite natural to study
the diffusive behaviour of the standard map~\cite{Lichtenberg}
which displays the typical features of low-dimensional Hamiltonian
systems, whose dynamics is given by
\begin{equation}
\left\{
{\begin{array}{lcl}
J_{t+1} & = & J_t + K \sin \phi_t\\
\phi_{t+1} & = & [\phi_t + J_{t+1}] \,\,\,\,\,\mbox{mod}\,\,\,\, 2\pi\,.
\end{array}
}\right .
\label{eq:stdmap}
\end{equation}
As soon as $K>0$ the system is not integrable and bounded chaotic
regions appear. However, in order to have a fully developed diffusion 
in the variable $J_t$, 
$K$ must be larger than $K_c \approx 0.97$. A very crude approximation,
amounting to assume $\phi_t$ as a sequence of independent variables,
gives
\begin{equation}
 D_J^e \simeq {K^2 \over 2} \langle{\sin^2 \phi}\rangle = 
{K^2 \over 4}\,.
\label{eq:diffstdmap}
\end{equation}
A more precise argument~\cite{Rechester}, that is able to take into
account the time correlation of $\phi_t$ is in good agreement with
$D_J^e$ values obtained by numerical computation for $K > 3$ with the
exception of small regions around some specific values of $K$.
Anomalous diffusion occurs for such specific values, where
ballistic-like trajectories of $J_t$ appear, known as accelerator
modes, whose regular dynamics gives a non integrable contribution to
the velocity-velocity correlation function and produces
superdiffusion~\cite{Zaslavsky}, \cite{Ishizaki}, \cite{Robnik}.  For
instance at $K=6.9115$, one observes $\nu \simeq 0.66$ and a non
Gaussian shape of the $P(J,t)$~\cite{Castiglione}.  However, even for
$K$-values admitting accelerator modes, the addition of a small noise
to the dynamics makes the diffusion normal~\cite{Majda}.  The noise
term, indeed, avoids the stickiness to the accelerator modes.

%>>>>>>>>>>>>>>>>>>>>>>>>>>>>>>>>>>>>>>>>>>>>>>>>>>>>>>>>>>>>>>>>>>>>
\subsubsection{Pair dispersion in fully developed turbulence}
\label{sec:2.1.3}
%>>>>>>>>>>>>>>>>>>>>>>>>>>>>>>>>>>>>>>>>>>>>>>>>>>>>>>>>>>>>>>>>>>>>
So far, we have considered the behaviour of a single 
particle, here we show that anomalous diffusion naturally arises even when 
studying the two-particles properties of a fluid. 
Consider indeed the dispersion of a pair of particles advected by a
homogeneous, isotropic, fully developed turbulent field. Such a problem
is rather important in applications as it describes 
the spreading of pollutants in presence of turbulence.

Because of the incompressibility of the velocity field, the distance
between the two particles will, on average, increases~\cite{Cocke}.
Richardson~\cite{Richardson} proposed for the PDF, $P({\mathbf r},t)$, 
of the two-particle separation $\mathbf r = \mathbf r_2 - \mathbf r_1$, the 
following evolution equation in $d$ dimensions
\begin{equation}
   \partial_t P({\mathbf r},t) = 
       \left[\nabla\cdot(K(r)\nabla)\right]P({\mathbf r},t)\,.
   \label{eq:richardson}
\end{equation}
The effects of turbulence are taken into account 
in the turbulent diffusivity $K(r)$. 
From a collection of experimental data, Richardson
\cite{Richardson} proposed his ``$4/3$'' law
$$
K(r) = \alpha r^{4/3}\;.
$$
It is easy to realize a posteriori that the above relation is a 
straightforward consequence of the Kolmogorov scaling~\cite{Frisch}. 
Assuming isotropy, Eq.~\eqref{eq:richardson} can be written as 
\begin{equation}
   \partial_t P(r,t) = {1 \over r^{d-1}} {\partial \over \partial r}
       \left (K(r) r^{d-1} {\partial \over \partial r} P(r,t) \right )\,,
   \label{eq:richardsonr}
\end{equation}
whose asymptotic solution is the non Gaussian PDF:
\begin{equation}
   P(r,t) \sim \frac{1}{t^{3/2}} \exp 
      \left [ { -C \left ( {r \over t^{3/2}} \right )^{2/3} } \right ]\,\,.
   \label{eq:pdfdiff}
\end{equation}
Thus, the growth law for the average separation between the particles is
\begin{equation}
   \langle{|r(t)|^q}\rangle \sim t^{\nu q}\,,
   \label{eq:turbodiff}
\end{equation}
with $\nu=3/2$.
In spite of its apparent simplicity, Richardson's approach is able to
describe, in a quantitative way, the pair dispersion in synthetic
turbulent fields~\cite{Boffetta} and in turbulent fields obtained by
direct numerical simulations as well~\cite{Boffetta2}.  

Moreover, Eq.~\eqref{eq:richardsonr} can be used to describe a rather
general class of transport processes. As an example, we can mention the 
diffusion on fractal geometries of dimension $d_f = d$, where a
radius-dependent diffusion coefficient $K(r) \propto r^{\xi}$ 
can be assumed~\cite{Procaccia1985}.

The general solution corresponding to Eq.~(\ref{eq:richardsonr})
with $K(r) \propto r^{\xi}$ is:
\begin{equation}
   P(r,t) \sim \frac{1}{t^{\frac{1}{2 - \xi}}} 
           \exp \left[-C \left ({r \over t^{1 \over 2-\xi}} 
                         \right )^{2-\xi} \right ]\,\,,
   \label{eq:pdfdiffgeneral}
\end{equation}
which extends formula~\eqref{eq:pdfdiff}.
Therefore the ``typical position'' is $X_f \sim t^{1 \over {2-\xi}}$, 
and, changing $\xi$, we can explore subdiffusive ($\xi<0$), normally
diffusive ($\xi=0$) and superdiffusive ($\xi>0$) regimes.

%>>>>>>>>>>>>>>>>>>>>>>>>>>>>>>>>>>>>>>>>>>>>>>>>>>>>>>>>>>>>>>>>>>>>>>
\subsection{Strong anomalous diffusion in stochastic processes}
\label{sec:2.2}
%>>>>>>>>>>>>>>>>>>>>>>>>>>>>>>>>>>>>>>>>>>>>>>>>>>>>>>>>>>>>>>>>>>>>>>
As we have seen before, in the presence of anomalous diffusion
the PDF of the process is typically no longer Gaussian.  A natural
issue is to determine the shape of $P(x,t)$, at least for large
times. A simple assumption, looking at Eqs.~\eqref{eq:richardsonr}
and~\eqref{eq:pdfrsf}, is the scaling behaviour
\begin{equation}
   P(x,t) \; \simeq \; t^{-\nu} \, F \left ( {|x| \over t^\nu} \right )\,\,.
\label{eq:scaling}
\end{equation}
The normal diffusion corresponds to $\nu = 1/2$ and $F(z) = 
\exp(-\mbox{c} z^2)$. 
For $\nu\ne 1/2$, it has been proposed~\cite{Bouchaud}
\begin{equation}
F(z) \propto e^{-c|z|^{\alpha}}\,\,.
\label{eq:effez}
\end{equation}
One could hope that $\alpha$ is a function of $\nu$ only, but
in general this is not true. Nevertheless, it is interesting
to note that both Eqs.~(\ref{eq:scaling}) and (\ref{eq:effez})
are consistent with an argument \'a la Flory due to Fisher in the context 
of polymer physics \cite{MEFisher} predicting:
\begin{equation}
P(x,t) \sim t^{-\nu} \exp \left [ -c 
        \left ( {|x| \over t^\nu} \right )^{1 \over 1 - \nu} 
                          \right ]\,\,,
\label{eq:pdfflory}
\end{equation}
with $\alpha=1/(1 - \nu)$.  We will see in the following that the
shape~(\ref{eq:pdfflory}) has an interesting consequence for the front
propagation in reactive systems, since transport processes whose PDF
scales like (\ref{eq:pdfflory}) give rise to a reactive-transport with
constant propagation velocity.  In general, when a reactive term is
added to a transport-diffusion equation not only $\nu$ but also the
exponent $\alpha$ is important to determine the front propagation
features.

Let us remark that the scaling \eqref{eq:scaling} implies 
\begin{equation}
\langle |x(t)|^q \rangle \sim t^{q\nu},
\label{eq:mscale}
\end{equation}
therefore, a single exponent is able to characterize the asymptotic behaviour 
of all the moments. We refer to this case as {\it weak} anomalous diffusion.
However, certain deterministic and stochastic processes~\cite{Klages} 
show that the simple scaling~\eqref{eq:mscale} is not always verified, and a 
more intricate scenario characterized by the generalized scaling 
\begin{equation}
\langle |x(t)|^q \rangle \sim t^{q\nu(q)}
\label{eq:gen_scale}
\end{equation}
instead holds, where also the exponent $\nu = \nu(q)$ depends on $q$. 
This property is referred to as 
{\em strong anomalous diffusion}~\cite{Castiglione}.

Remarkably strong anomalous diffusion is also present in apparently 
simple stochastic processes like some continuous time random walks 
(CTRW)~\cite{Klafter}, that are a variant of the classic random walk (RW),
where a particle undergoes a series of kicks (collisions) at random times
$t_1,t_2,\ldots,t_n,\ldots$ and between two consecutive collisions the 
velocity remains constant.
The position of the particle at time $t$, such that $t_{n} < t \leq t_{n+1}$,
will be
\begin{equation}
x(t) = x(t_{n}) + v_n(t-t_{n})
\label{eq:CTRW}
\end{equation}
and both the time intervals $\tau_n = t_{n+1} - t_{n}$ and the 
velocities $v_n$ are independent random variables with arbitrary
distribution~\cite{Klafter}. In the 
following we consider a variant with a truncated power law distribution
\begin{equation}
P(\tau) \propto 
\begin{cases}
\tau^{-g} & \quad  1 \leq \tau \leq T \\
0         & \quad  \mbox{elsewhere}
\end{cases}
\label{eq:cases}
\end{equation}
with $g > 1$ and $v_{n} = \pm 1$ with equal probability.
The presence of the cutoff $T$ implies that the hypothesis 
of the CLT for the process~(\ref{eq:CTRW}) are
fulfilled, thus as $t\gg T$, it converges to a Gaussian process.
However for $T$ sufficiently large \cite{TrunkLevy}, the
convergence can be made so slow that a long and robust
pre-asymptotic regime of strong anomalous diffusion takes place.

A quantity that will be important in the following is the
$q$-order moment of the waiting time $\tau$ whose asymptotic
scaling for large $T$ is the following
\begin{equation}
\langle\tau^q\rangle_c \sim  
\begin{cases} 
T^{1-g+q}    & \mbox{if~} q > g-1 \\
a(q,g)       & \mbox{if~} q < g-1 
\end{cases} 
\label{eq:tau_q}
\end{equation}
where $a(q,g)$ is a constant independent of $T$ and
the index $c$ indicates the average over the ``truncated''
distribution.
At $T\to \infty$, various diffusive regimes occur depending on the
value of the exponent $g$, see~\cite{Andersen} and~\cite{Forte}, 
in particular the case $g \in (3,4]$
corresponds to the anomalous diffusion, $\langle |x(t)|^{q}\rangle\sim
t^{q\nu(q)}$, with
\begin{equation}
\label{eq:g_scaling}
q\nu(q)=
\begin{cases} 
q/2  , & q=1,2 \\ 
q+2-g, & q=3,4,5,\cdots
\end{cases}
\end{equation}
The above behaviour of $q\nu(q)$ is quite peculiar as ,for $q\leq 2$, 
it corresponds to normal diffusion, while, for larger
$q$, we have $q\nu(q) \ne q/2$, and this represents an example of
strong anomalous regime.

\begin{figure}[h!]
\mbox{ }\vspace{0.5cm}\\
\begin{center}
\includegraphics[width=0.4\textwidth]{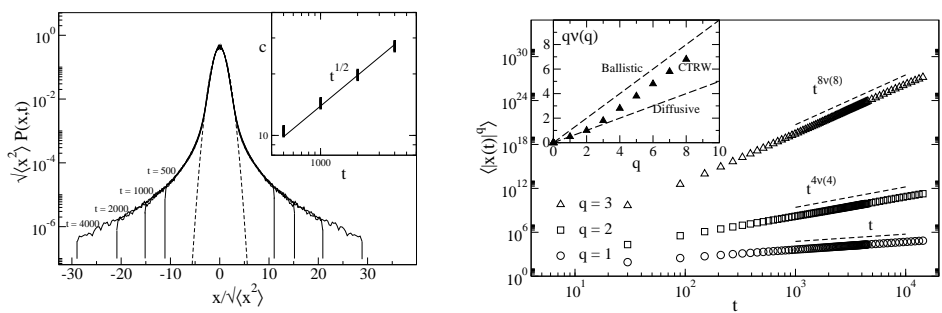}
\hspace{0.5cm}
\includegraphics[width=0.45\textwidth]{Fig_1b.eps}
\caption{Statistical features of the CTRW process~\eqref{eq:CTRW}
and~\eqref{eq:cases}.
On the left, rescaled probability densities at different times for
$g=3.2$.  The vertical lines mark the truncated probability
distribution at the bounded domain, $|x|\le c(t)$, the dashed lines
represents the Gaussian.  The inset shows the scaling
$c(t)/\sqrt{t}\sim t^{1/2}$ of the domain size.  On the right, power
law scaling of the moments with time, the inset reports the
nonlinearity of $q\nu(q)$ as it defined in Eq.~\eqref{eq:g_scaling}.
Adapted from~\cite{Forte}.}
\label{fig:Dist_CTRW} 
\end{center}
\end{figure}

The above behaviour of $q\nu(q)$ is quite peculiar as it coincides
with that one of normal diffusion for $q\leq 2$, while for larger $q$,
we have $q\nu(q) \ne q/2$, and this represents an example of strong
anomalous regime.  The relation (\ref{eq:g_scaling}) provides strong
indications on the possible form of the PDF bulk; the lowest order
moments behave in time as in the case of normal diffusion, thus we
expect that the $P(x,t)$ has a Gaussian-like bulk which scales as
$P(x,t)=t^{-1/2} f(xt^{-1/2})$) for moderate values of the argument
$|x/\sqrt{t}|$.  The collapse of the rescaled PDF of the
CTRW~(\ref{eq:CTRW}) at different times is shown in
Figure~\ref{fig:Dist_CTRW}. \\
It is worth remarking that the above
result is not actually a violation of the CLT, as in the bulk all the
PDFs collapse onto a Gaussian and only the tails deviate from it.  The
CLT, indeed, does not grant anything on the nature of the tails, it
only specifies the shape of the limit distribution within the scaling
region, $|x(t)/\sqrt{t}| \sim O(1)$. The tails outside such a region
are not universal and generally not Gaussian.

\vspace{0.2truecm}

%------------------------------- FIG.1 -------------------------------

%>>>>>>>>>>>>>>>>>>>>>>>>>>>>>>>>>>>>>>>>>>>>>>>>>>>>>>>>>>>>>>>
\subsection{Diffusion on graphs}
\label{sec:diffgraph}
%>>>>>>>>>>>>>>>>>>>>>>>>>>>>>>>>>>>>>>>>>>>>>>>>>>>>>>>>>>>>>>>
In this section we briefly discuss a transport process on the discrete
structure of graphs. An undirected graph is a collection of vertices
pairwise connected, or not, by links~\cite{bollobas1998}.  An $N\times
N$ matrix, $A$, called adjacency matrix, can be associated to each
graph of $N$-vertices's, with entries $A_{ij}$ = $1$ if there is a link
between vertices's $i$ and $j$, $A_{ij}$ = $0$ otherwise. A fundamental
quantity is the degree of nodes, $k_i = \sum_j A_{ij}$, that measure
the number of links established by each node $i \in [1..N]$.

An unbiased random walk on a graph can be defined whenever 
a walker at time $t$ on the node $i$ can jump at time $t+1$
on the node $j$ with a transition probability $A_{ij}/k_i$.
The latter assumption of equal-probability
of the transition (possible only when $A_{ij}=1$)
to nearest neighbours of $i$ can be relaxed 
by introducing some bias in the jump probabilities.

The diffusion properties of a random walk on graphs depend on both the
{\em fractal dimension} \cite{Mandelbrot} $d_f$, that is related to
the scaling of the number of points in a sphere of radius $\ell$:
$N(\ell) \sim \ell^{d_f}$, and the {\em spectral dimension} $d_s$,
that is defined by the return probability $q_{t}(x)$ to a generic site
$x$ in $t$ steps $q_{t}(x) \sim t^{-d_s/2}$~\cite{Alexander_Orbach}.
In isotropic structures, the mean square displacement
scales as (see~\cite{BurioniCassi} and~\cite{BurioniCassi2})
\begin{equation} 
\langle x^{2}(t)\rangle\sim t^{d_s/d_f},
\label{eq:msd_graph}
\end{equation}
while, in anisotropic graphs, the above equation can fail
(see~\cite{Bertacchi} and \cite{Mendez}).
The continuous-time diffusion on graphs~\cite{bollobas1998} is ruled 
by the following master equation
\begin{equation}
{{\mathrm d} P_i \over {\mathrm d}t}=
D\sum_{j} W_{ij} P_j
\label{eq:graphdiffusion}
\end{equation}
where $W_{ij}= A_{ij} - k_i \delta_{ij}$ is the Laplacian on the
graph and $k_i$ is the degree of the node $i$.
Given the graph properties one can wonder about the behaviour of
high-order moments and the possible shape of the PDF.

\begin{figure}[ht!]
\centering
\includegraphics[width=6.2cm]{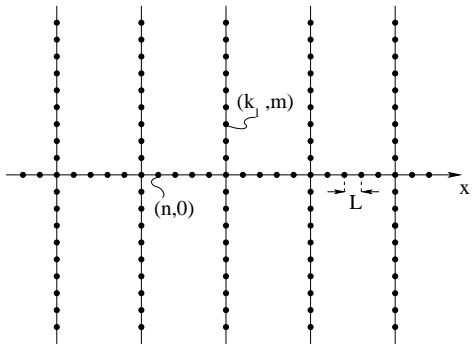}
\caption{Representation of a comb lattice, as an example of a graph
with a branched structure. Adapted from~\cite{mancinelli2}.
\label{fig:comb}
}
\end{figure}

Just to give an example of diffusion on graph, we consider
the comb lattice in Fig.~\ref{fig:comb}. 
The structure of the comb is such as sites $(n,0)$, where 
$n=0,\pm 1,\pm 2, \cdots$, define the backbone (horizontal line) 
of the graph, and the ``teeth'', i.e., sites $(k_l,m)$, 
where $m=0,\pm 1,\pm 2, \cdots$ and $k_l$ labels the x-coordinate
of the $l-$th ``tooth'' (in Fig.~\ref{fig:comb} $k_l=5l$), gives
the lateral escaping structures. 
%---------------------- FIG.2 -----------------------------------
%-----------------------------------------------------------------
Denoting by $P_{\mathbf i}(t)$ the probability to be on the site
${\mathbf i}=(n,m)$ at time $t$, the evolution equation is given
by~\eqref{eq:graphdiffusion}, where $A_{ij}$ is the adjacent matrix of
the comb.  Because of the possibility to ``escape'' along the vertical
teeth, the diffusion process on the horizontal direction has a
subdiffusive character~\cite{Cassi},\cite{Forte2}:
\begin{equation}
   \langle{|x(t)|^2}\rangle \sim t^{1/2} \qquad \mbox{i.e.}\;\;\;\;\;\nu=1/4\,.
   \label{eq:difcomb}
\end{equation}
Numerical simulations~\cite{mancinelli2} show that the PDF $P(x,t)$ 
behaves like
\begin{equation}
   P(x,t) \sim \frac{1}{t^{1/4}} \exp \left 
       [-C\,{\left ( |x| \over t^{1/4}\right )^{4/3}} \right ]\,\,,
   \label{eq:pdfcomb}
\end{equation}
and it is worth to note that in this case the Fisher's
conjecture~\eqref{eq:pdfflory} holds.

%>>>>>>>>>>>>>>>>>>>>>>>>>>>>>>>>>>>>>>>>>>>>>>>>>>>>>>>>>>>>>>>>>>>>
\subsubsection{Normal diffusion on fractal graphs}
\begin{figure}[h!]
\centering
\includegraphics[width=0.5\textwidth]{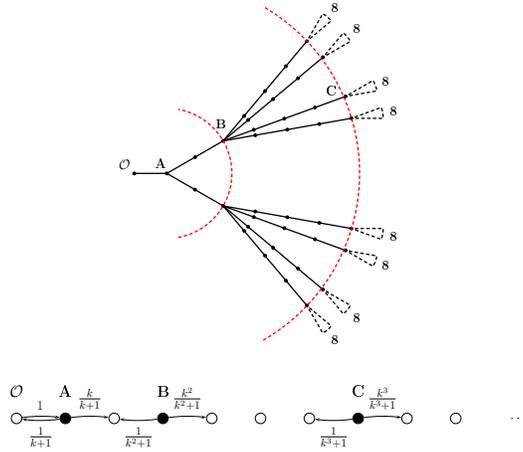}
\caption{Geometrical construction of a Nice Tree of dimension $k = 3$.
The origin $\mathcal{O}$ is connected with a point $A$ by a link of length 1;
from $A$ the tree splits in $k$ branches of length $2^{1}$ each.
The end point of such branches, in turn, splits again into $k$ branches of
length $2^{2}$ and so on. 
Below, the transition rules for a walker 
to be one step closer or farther to the origin $\mathcal{O}$, full dots
mark the branching points. Adapted from~\cite{Forte}.
\label{fig:NTD}
}
\end{figure}

%>>>>>>>>>>>>>>>>>>>>>>>>>>>>>>>>>>>>>>>>>>>>>>>>>>>>>>>>>>>>>>>>>>>>
Consider now the diffusion in a special class of
graphs the {\em Nice Trees of dimension} $k$ (NT$_{k}$) that are 
recursively defined from an origin $\mathcal{O}$ 
as it is described in Fig.~\ref{fig:NTD}.
%---------------------------------- FIG.3 ----------------------------
%-------------------------------------------------------------------
Such trees are characterized by the remarkable property that fractal and
spectral dimension coincide \cite{BurioniCassi} and \cite{BurioniCassi2},
\begin{equation}
d_{f} = d_{s}(k) = 1 + \frac{\ln{k}}{\ln{2}}
\label{eq:NTD_ds}
\end{equation}
therefore, despite the nontrivial structure, Eq.~(\ref{eq:msd_graph})
implies a normal behaviour $\langle x^{2}(t)\rangle\sim t$ for any
values of $k$.  However, the linear growth of the mean square
displacement in time (i.e. $2\nu(2)=1$) does not necessarily imply a
Gaussian diffusive behaviour.  Therefore to fully characterize the
diffusion properties of the unbiased random walk on the NT$_{k}$, we
need to determine the probability distribution that a walker lays at
distance $x$ from the origin $\mathcal{O}$. To this purpose, we assign
to the site the integer distance $x$ if it is connected to
$\mathcal{O}$ by a minimal path with $x$-links.  It is remarkable that
the projected process obtained by using the new variable $x$ still
maintains the Markov properties and it is ruled by the following
(discrete time) master equation~\cite{Forte}:
\begin{eqnarray}
P_{t+1}(x-1) =& \displaystyle{\frac{1}{2} P_t(x-2) + \frac{1}{k+1} P_{t}(x)} 
\nonumber \\
P_{t+1}(x)   =& \displaystyle{\frac{1}{2} P_t(x-1) + \frac{1}{2} P_{t}(x+1)} 
\nonumber \\
P_{t+1}(x+1) =& \displaystyle{\frac{k}{k+1} P_t(x) + \frac{1}{2} P_{t}(x+2)} 
\label{eq:1D_NTD}
\end{eqnarray}
where, with reference to Fig.~\ref{fig:NTD}, the first and third equations 
holds only for branching points ($x=2^n-1$), the second one for all the other sites.
Accordingly, the transition matrix $w(x\pm 1|x)$ of the one-dimensional RW
from $x \to  x\pm 1$, in a time step reads
\begin{equation}
 w(x+1|x) =
\begin{cases} 
\displaystyle{\frac{k}{k+1}},  &\mbox{if~}\; x=2^n-1\\ 
1/2,    &\mbox{elsewhere} 
\end{cases}  
\end{equation}
\begin{equation}
  w(x-1|x) = 
  \begin{cases}
 \displaystyle{\frac{1}{k+1}}, &\mbox{if~} x=2^n-1\\
 1/2,   &\mbox{elsewhere} 
\end{cases}
\end{equation}
where $2^n-1$ is the formula identifying the distance of the branching points
from the origin $\mathcal{O}$,
$W(1|0) = 1$ is the condition for a reflecting boundary in $\mathcal{O}$.
The RW on NT$_{k}$ is thus mapped onto a RW on a one-dimensional lattice
in a deterministic heterogeneous environment.
The physical interpretation of the transition rules are straightforward:
if a walker stays on a branching point, there are $k$ possibilities to
go one step away from origin and $1$ possibility to make one step closer.
Then the next step will take it either
farther from the origin with probability $p_{+}= k/(k+1)$ or closer the origin
with probability $p_{-}= 1/(k+1)$. Whereas if a walker is outside the
branching point, both steps are unbiased, $p_{-} = p_{-} = 1/2$.
The inhomogeneity stems from the branching points $x=2^{n}-1,\;n=1,2,3\cdots$ 
which represent special points (``defects'') but become exponentially rare 
as long as the walker lies far away from the origin. Thus, far away from 
the origin, the process remains an unbiased RW for so long time that 
``Gaussian character'' of the distribution is not altered by the presence 
of defects.

\begin{figure}[h!]
\centering
\includegraphics[width=0.5\textwidth]{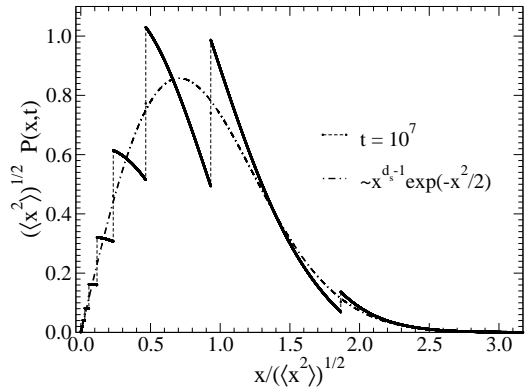}
\caption{Probability density obtained by iterating the master equation
\eqref{eq:1D_NTD} for a $NT_k$ with $k=2$. Dashed line indicates 
the interpolation with the Gaussian radial distribution in dimension 
$d_s$. Adapted from~\cite{Forte}.
\label{fig:Dist}
}
\end{figure}

Figure~\ref{fig:Dist} shows the simulation results for the
probability density $P_{t}(x)$
rescaled to $x\to x/\sqrt{\langle x^{2}\rangle}$ and
$P_{t}(x)\to\sqrt{\langle x^{2}\rangle}P_{t}(x)$ obtained by iterating
the master equation~\eqref{eq:1D_NTD} for NT$_k$ with $k = 2$.
The dash-dotted curve is the approximated solution
\begin{equation}
F_t(x) = 
\frac{2 x^{d_s-1}}{\Gamma(d_s/2)(2t)^{d_s/2}} \exp(-x^{2}/2t)
\label{eq:app_NTDpdf}
\end{equation}
which well interpolates the exact numerical result.
Expression~\eqref{eq:app_NTDpdf} is the radial Gaussian
distribution in dimension $d_s(k)$ 

%---------------------------------- FIG.4 ----------------------------
%-------------------------------------------------------------------
As a final remark we note that, despite the geometrical complexity of
NTk, the large scale statistical properties of RW on this graph
remains Gaussian-like, as, on the other hand, it expected by the shape
of the approximated distribution
\eqref{eq:app_NTDpdf}.

%%%%%%%%%%%%%%%%%%%%%%%%%%%%%%%%%%%%%%%%%%%%%%%%%%%%%%%%%%%%%%%%%%%%%
\section{Front propagation in reactive transport systems 
with anomalous diffusion}
\label{sec:front_Adiff}
%%%%%%%%%%%%%%%%%%%%%%%%%%%%%%%%%%%%%%%%%%%%%%%%%%%%%%%%%%%%%%%%%%%%%
In the previous Section, different mechanisms for anomalous
diffusion, corresponding to suitable transport operators $\hat{L}$, 
have been discussed. Here we consider systems where 
anomalous diffusion is coupled to a reaction dynamics leading to a
generalized reaction diffusion equation
\begin{equation}
{\partial \theta \over \partial t}=\hat{L} \theta 
+{1 \over \tau} f(\theta)\, 
\label{eq:genFKPP}
\end{equation}
whose solution can be compared with the classical theoretical 
framework corresponding to the FKPP Equation \eqref{eq:FKPP}.
In particular, we focus on the asymptotic features of front
propagation generated by Eq.~(\ref{eq:genFKPP}).
\\
We recall that the FKPP equation admits propagating 
fronts separating regions with $\theta \simeq 1$ and 
$\theta \simeq 0$. The standard propagation scenario is then defined by a 
constant front speed $v_f=2 \sqrt{D_0 f^{'}(0)/\tau}$
(see \cite{Fisher} and \cite{KPP}) with a thickness of the front given by
$\xi = 8\sqrt{D \tau/f^{'}(0)}$ and an exponential 
decay of $\theta$ at large x
\begin{equation}
   \theta(x,t) = F(x-v_f t) \qquad \mbox{with} \qquad
    F(z) \sim e^{-z/\xi} \,\,\,\,\,\,\mbox{for}\,\,\,\,\, z \gg 1\,.
   \label{eq:FKPPscen}
\end{equation}
 
When dealing with Eq.\eqref{eq:genFKPP} a question naturally arises
about the conditions needed to change the FKPP propagation scenario.
We will see that if the transport operator $\hat{L}$ at large time
exhibits a normal diffusion behaviour and the reaction time $\tau$ is
large enough, then a standard propagation scenario establishes. However,
the presence of anomalous diffusion alone it is not sufficient 
to guarantee an anomalous transport~\cite{Volpert}, i.e. a non 
constant front speed.  We will see that
for the reaction spreading is relevant not only the anomalous exponent
$\nu$ but the details of the mechanism play a major role.  This can be
explicitly shown for some cases, see~\cite{mancinelli}
and~\cite{mancinelli2} for details.

Studying different reacting systems in the presence of anomalous 
diffusion, we discuss 3 different scenarios:
\begin{description}
   \item [a)] exponential decay of $\theta$ with 
              constant front speed $v_f$ and thickness $\xi$,
              i.e. Eq.~(\ref{eq:FKPPscen});
   \item [b)] exponential decay of $\theta$ but $v_f$ and $\xi$
              can vary in time as a power law;
   \item [c)] power law decay of $\theta({\mathbf x},t) \sim |x-v_f t|^
              {-\alpha}$ and $v_f$ exponentially increases in time.
\end{description}

%>>>>>>>>>>>>>>>>>>>>>>>>>>>>>>>>>>>>>>>>>>>>>>>>>>>>>>>>>>>>>>>>>>>>
\subsection{Some general results}
%>>>>>>>>>>>>>>>>>>>>>>>>>>>>>>>>>>>>>>>>>>>>>>>>>>>>>>>>>>>>>>>>>>>>
Consider the case in which the generalized diffusion
equation~\eqref{eq:passive} generates asymptotically normal diffusion.
It is possible to prove \cite{Freidlin} a relation between the solution of 
Eq.~\eqref{eq:genFKPP} and the Lagrangian trajectories ${\bf x}(t)$ 
associated to the diffusion process~\eqref{eq:passive}:
\begin{equation}
   \theta({\bf x}, t) = \left \langle { \theta({\bf x}(0), 0)\,
   \exp \left ( {1 \over \tau} \int_0^t {f(\theta({\bf x}(s;t),s))
   \over \theta({\bf x}(s;t),s)} {\rm d} s \right ) } \right
   \rangle\,,
   \label{eq:feykacex}
\end{equation}
where the average is performed over all the trajectories ${\bf
x}(s;t)$ that started in ${\bf x}(0)$ and ended in ${\bf x}(t;t) =
{\bf x}$.  Equation~(\ref{eq:feykacex}), although rather important
from a theoretical point of view, cannot be easily used for explicit
computations. However it suggests the implementation of an efficient
stochastic algorithm for numerical simulation of the
Eq.~\eqref{eq:genFKPP} (see~\cite{Abel}).  In addition, using the maximum
principle~\cite{Freidlin} and noting that, because of the convexity of
$f(\theta)$, $f(\theta)/\theta \leq f'(0)$ one can write an upper
bound for $\theta$ in terms of the solution of the linearized
Eq.~\eqref{eq:genFKPP}:
\begin{equation}
\partial_t \theta_L = \hat{L} \theta_L + \frac{f'(0)}{\tau} \theta_L\,.
\label{eq:lard}
\end{equation}
In fact, if $\theta(\mathbf x,0) \leq \theta_L(\mathbf x,0)$ the maximum
principle~\cite{Freidlin} implies
\begin{equation}
   \theta(\mathbf x,t) \leq \theta_L(\mathbf x,t) \,
   \label{eq:iniz}
\end{equation}
for all times.
From Eq.s~(\ref{eq:feykacex}-\ref{eq:iniz}) one obtains
\begin{equation}
   \theta({\bf x},t) \leq \theta_L({\bf x},t) = 
\langle \theta({\bf x}(0;t),0)\rangle \exp \left (
                      {f'(0)\over \tau} t \right ) \,,
   \label{eq:upbound}
\end{equation}
where $\langle \theta({\bf x}(0;t),0)\rangle=P({\mathbf x},t)$
is the solution of Eq.~\eqref{eq:passive} with initial condition $\theta({\bf
x},0)$ (that we assume localized around ${\bf x}= 0$).  
In the Introduction and in Sect.~\ref{sec:diffusion}, we saw that under 
general conditions (i.e., spatial and temporal short-range correlations) 
Eq.~\eqref{eq:passive} has the same 
asymptotic behaviour of a Fick equation.  As a consequence
(see also Eq.~\eqref{eq:pdfFick}), we have
\begin{equation}
\langle \theta({\bf x}(0;t),0)\rangle \sim 
  \exp \left\{-\frac{1}{4t} \sum_{i,j} 
x_i \left[D^e\right]^{-1}_{ij} x_j \right\}\,.
   \label{eq:apgauss}
\end{equation}
Eq.s (\ref{eq:upbound}) and (\ref{eq:apgauss}) imply that,
along the $x$-direction, the field $\theta$ is exponentially small
until a time $t$ of the order of $x/\sqrt{4 D^e_{11} f'(0)/\tau}$.
Therefore, considering just 1D propagation along the x direction,
we have an upper bound for $v_f$
\begin{equation}
   v_f \leq 2 \sqrt{D^e_{11} f'(0)/\tau}\,.
   \label{eq:vfupb}
\end{equation}
The above discussion shows that, if normal diffusion holds,
there is a front propagating with a constant finite speed.
Nevertheless, the analytical determination of $v_f$ is rather 
difficult even for simple laminar fields~\cite{Abel}.

%%%%%%%%%%%%%%%%%%%%%%%%%%%%%%%%%%%%%%%%%%%%%%%%%%%%%%%%%%%%%%%%%%%%%%
\subsection{Behaviour of the front position
\label{sec:3.3}}
\begin{figure}[ht!]
{\centerline{
\includegraphics[width=7cm]{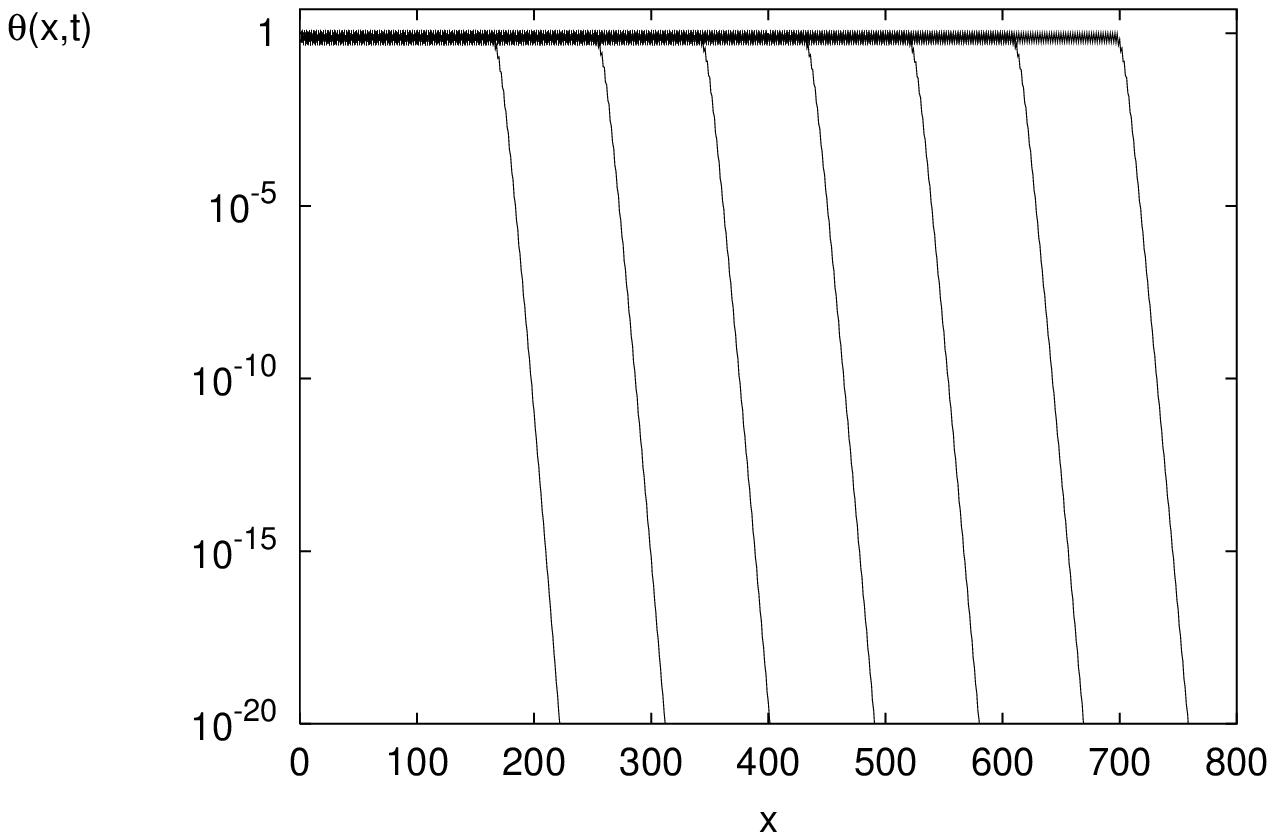}
\includegraphics[width=7cm]{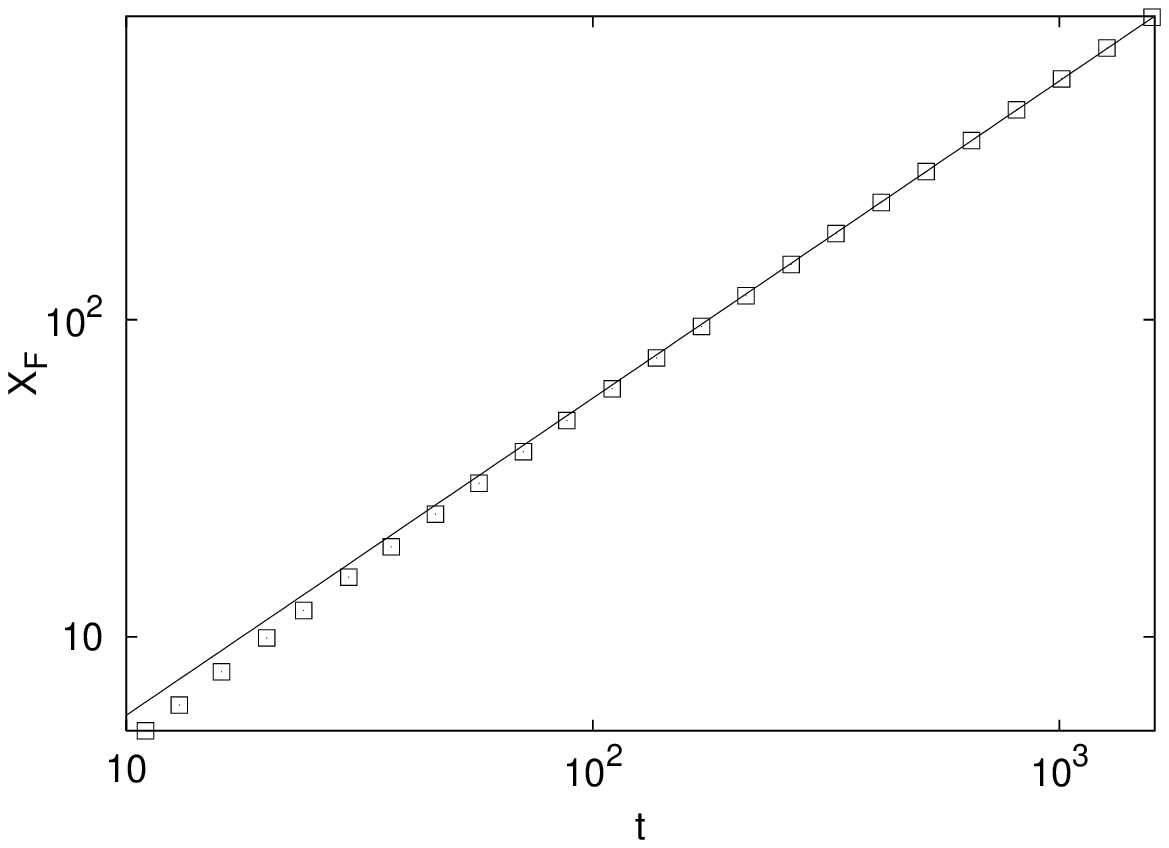}}}
\caption{Front propagation (on the left) and typical 
front position $X_f$ (on the right) for the
reaction dynamics on a comb lattice. The solid line
displays a linear behaviour. Adapted from~\cite{mancinelli2}.}
\label{fig:comb_front}
\end{figure}

%%%%%%%%%%%%%%%%%%%%%%%%%%%%%%%%%%%%%%%%%%%%%%%%%%%%%%%%%%%%%%%%%%%%%%
Let us now give an argument for the front propagation
in systems with anomalous diffusion.
As previously discussed a simple linear analysis,
using the Feynman-Kac like formula~(\ref{eq:feykacex}) 
and the maximum principle, leads to
Eq.~(\ref{eq:upbound}), here rewritten as
$$\theta(x,t)\le P(x,t)\, e^{\,c t},$$
where $c=f'(0)/\tau$.
When the scaling law~(\ref{eq:scaling}) together with
Eq.~(\ref{eq:effez}) holds, the following relation applies
\begin{equation}
  \theta(x,t) \le
      t^{-\nu} \, F \left ({|x| \over t^\nu} \right ) \, e^{c t} 
      \,\sim \,
      t^{-\nu} \, \exp \left [-C \left ({|x| \over t^\nu}\right )^\alpha 
           + c t \right ]\,\,.
  \label{eq:upbg}
\end{equation}
Imposing $\theta(X_f(t), t) \sim 1$, where $X_f(t)$ is the front
position, it follows from the previous bound
$\theta \lesssim e^{-C{(X_f t^{-\nu})}^\alpha} \; e^{c t} \sim 1$ and
therefore one has
\begin{equation}
X_f(t) \sim t^{\gamma} \; \; \; \;  
  \mbox{with} \; \; \;  \gamma=\nu+\frac{1}{\alpha}\,.
\label{eq:ubvar}
\end{equation}
In all numerical computations we always observe $X_f \sim
t^{\gamma}$, since for convex reaction terms the front is pulled by
perturbations of the unstable state, whose evolution is given by the
linearized problem.  Note that $\gamma$ depends not only on $\nu$ but
also on $\alpha$. However, if the Fisher's argument of
Eq.~(\ref{eq:pdfflory}) holds, i.e. 
if $\alpha = 1/(1-\nu)$, we have $\gamma=1$. That is,
despite the presence of anomalous diffusion, the FKPP
front propagation scenario remains valid. For example we can
consider two different cases, one of subdiffusion (comb lattice)
and another of superdiffusion (random shear flow) both in the
Flory class and, in both, the front propagation is standard.
Fig.~\ref{fig:comb_front} shows front propagation in the 
comb lattice.
%-------------------------- FIG.5 -------------------------------------
%----------------------------------------------------------------------

In order to find the shape of the front, we study Eq.~(\ref{eq:upbg})
around the typical position $X_f\sim t^{\gamma}$.  In fact,
introducing the displacement $\eta=x-X_f$, it results
$\theta\sim\exp{\left[ -\alpha C t^{-(\nu \alpha-\gamma(\alpha-1))}
\eta + o(\eta^2) \right ]}$. As $\gamma = \nu + 1/\alpha$
(see Eq.~(\ref{eq:ubvar})), 
$\theta\sim\exp{\left[ -\alpha C t^{1-\gamma}\eta + o(\eta^2) \right ]}$.
\\
Summarizing, the typical position of the front grows 
as $X_f \sim t^\gamma$~(\ref{eq:ubvar}) 
and the shape of $\theta(x,t)$ around $X_f$ is
\begin{equation}
   \theta(x,t) \sim  
      \exp \left[-{ {x-X_f(t) \over \xi(t)} } \right ] \;\;\;\;
      \mbox{with} \; \; \;  \xi(t) \sim t^\delta \;\;\;\;
         \mbox{and} \; \; \;  \delta=\gamma-1\,\,.
   \label{eq:thickvar}
\end{equation}
This implies, for instance, that in all the cases in which we observe
linear propagation of the front ($\gamma=1$), we also have a constant
front thickness ($\delta=0$), just as in the FKPP equation.  In
general, the above scenario remains valid also in the presence of
other anomalous diffusion processes providing that $P(x,t)$ be of the
form Eq.~(\ref{eq:scaling}) together with Eq.~(\ref{eq:effez}).
Conversely, if Eq.~(\ref{eq:scaling}) or Eq.~(\ref{eq:effez}) is no
more satisfied, and different propagation behaviour are allowed
(see~\cite{mancinelli} and~\cite{mancinelli2}).

%%%%%%%%%%%%%%%%%%%%%%%%%%%%%%%%%%%%%%%%%%%%%%%%%%%%%%%%%%%%%%%%%%%%%%
\subsubsection{Generalized Richardson diffusion}
\label{sec:3.2.1}
%%%%%%%%%%%%%%%%%%%%%%%%%%%%%%%%%%%%%%%%%%%%%%%%%%%%%%%%%%%%%%%%%%%%%%
It is instructive to apply the above analysis to a system whose transport 
process is given by the Richardson diffusion equation (as in
Eq.~(\ref{eq:richardsonr})), with the asymptotic solution 
given by~\eqref{eq:pdfdiff}. It is important to note that this
is not the case of a Flory shape (\ref{eq:pdfflory}).
We also recall that the growth law for the average separation 
between the particles is $\langle{|r(t)|^2}\rangle \sim t^{3}$.

The simplest way to study the reaction transport problem for fully
developed 3D turbulence is to consider Eq.~\eqref{eq:genFKPP} with the
transport operator ${\hat L}$ given by Eq.~\eqref{eq:richardsonr},
i.e.:
\begin{equation}
   \partial_t \theta(r,t) = {1 \over r^{d-1}} {\partial \over \partial r}
       \left (K(r) r^{d-1} {\partial \over \partial r} \theta(r,t) \right )
        + {1\over \tau} f(\theta)\,.
   \label{eq:rdrich}
\end{equation}
In Eq.~(\ref{eq:rdrich}) $r=0$ indicates the center of the spot.
One can wonder about the validity of (\ref{eq:rdrich}) as a reasonable
model for reaction process in turbulent field. It is known that, 
in systems with normal diffusion, for slow reaction,
i.e., $\tau \gg L/U$, the large scale and long time asymptotic
of the reaction advection diffusion equation
$\partial_t \theta + {\bf u} \cdot {\bf \nabla} \theta =
D \Delta \theta + f(\theta)/\tau$ is well approximated by a 
reaction diffusion equation with eddy diffusivity tensor 
$D^e$ of the corresponding problem~(\ref{eq:FP}), 
i.e., $v_f \simeq 2 \sqrt{D^e_{11} f'(0)/\tau}$~\cite{Abel}.
On the base of this result one can expect that 
Eq.~(\ref{eq:rdrich}) is a suitable approximation if $\tau$
is larger than the eddy turnover times in the inertial range.
This condition, hardly satisfied in chemical context,
has its relevance in ecological problems where
equations similar to~(\ref{eq:rdrich}) are considered~\cite{Okubo}.

%---------------------------- FIG.6 ------------------------------------
\begin{figure}
\centerline{
\includegraphics[scale=0.5,draft=false]{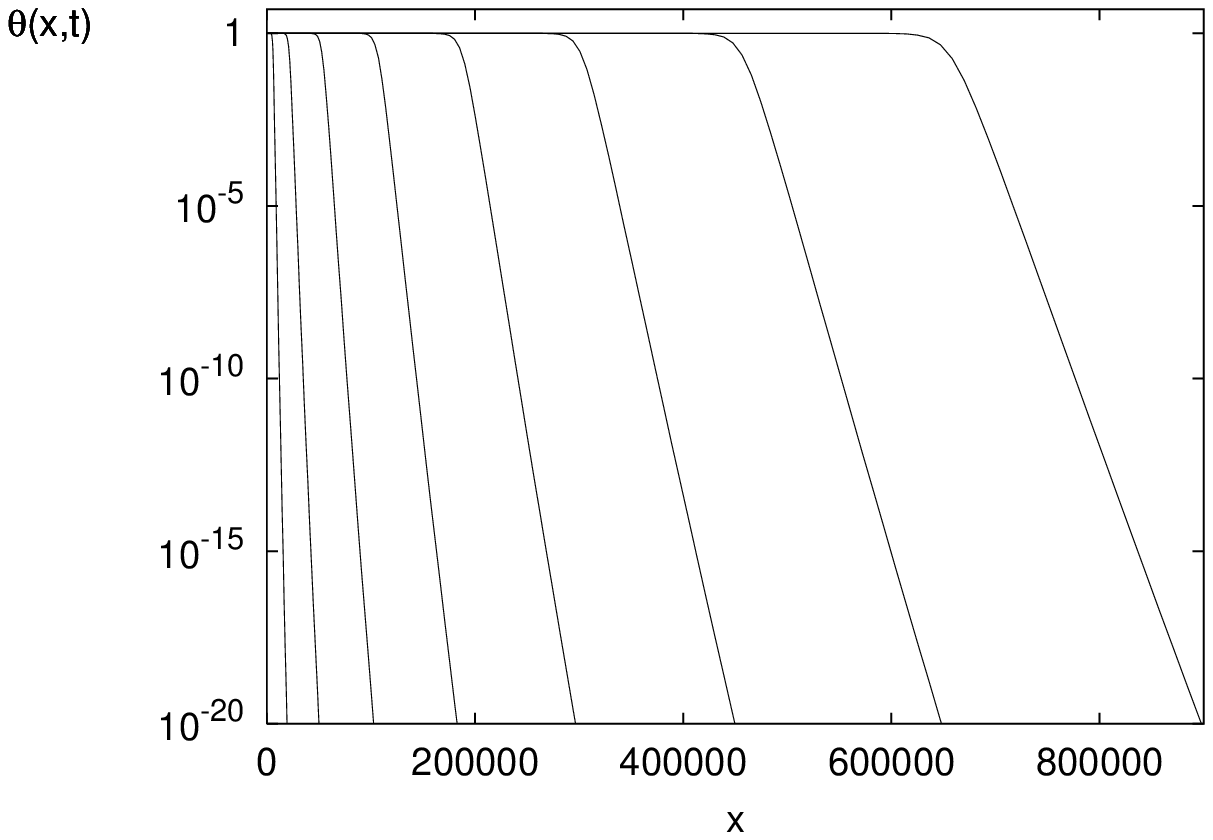}
\includegraphics[scale=0.5,draft=false]{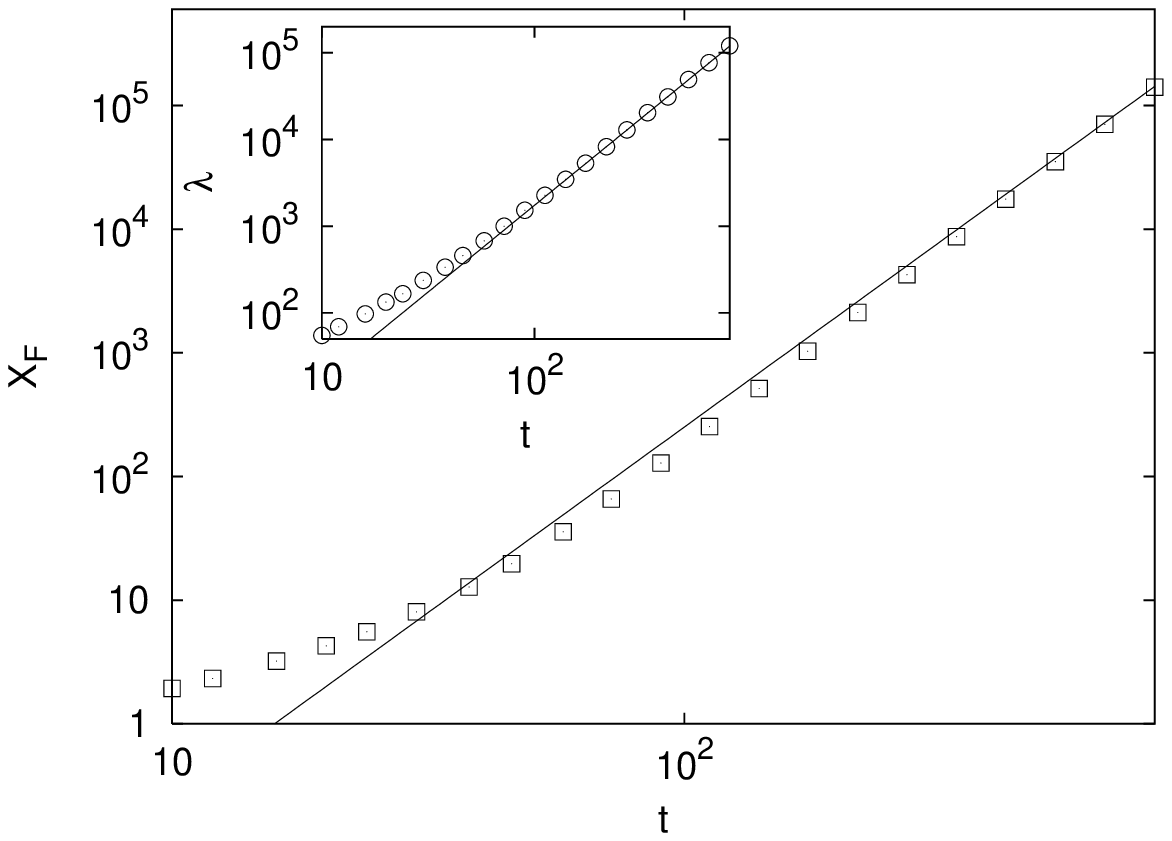}
}
\caption{Propagation features of reactive fronts with
Richardson diffusion in the case $K(r)=r^{4/3}$, 
$f(\theta)=\theta(1-\theta)$ and $\tau = 1$.
On the left, front shape at different times.
On the right typical position of the front $X_f$ vs $t$. 
The full line indicates the theoretical
prediction Eq.~(\ref{eq:ubvar}) with $\nu = 3/2$ and $\alpha=2/3$. 
The inset shows the front thickness $\xi$ together its theoretical 
prediction Eq.~\eqref{eq:thickvar}. Adapted from~\cite{mancinelli2}.
} 
\label{fig:richard}
\end{figure}
%------------------------------------------------------------------
Figure~\ref{fig:richard} shows the propagation and the typical position 
of the front.  It is apparent the good agreement between numerical 
results and theoretical prediction of Eq.~\eqref{eq:ubvar}.
In the inset of the right panel of Fig.~\ref{fig:richard}
the front thickness appears to be time dependent according to 
the prediction of Eq.~(\ref{eq:thickvar}),
i.e., the front shape changes as the front propagates. We remind that
the standard propagation is characterized by a rigid translation of
the front (e.g., see Eq.~(\ref{eq:FKPPscen})).

%>>>>>>>>>>>>>>>>>>>>>>>>>>>>>>>>>>>>>>>>>>>>>>>>>>>>>>>>>>>>>>>>>>>>>>
\subsubsection{Diffusion processes with power law tails}
\label{sec:3.2.3}
%>>>>>>>>>>>>>>>>>>>>>>>>>>>>>>>>>>>>>>>>>>>>>>>>>>>>>>>>>>>>>>>>>>>>>>
It is also interesting to investigate reaction processes
in the case of diffusion dynamics with power law tails. 
As an example we consider the L\'evy flight which follows
the rule
\begin{equation}
x(t+h)=x(t)+ h v(t)
\label{eq:levy}
\end{equation}
and $v(t)$'s are independent identically distributed (i.i.d.) 
random variables with a PDF decaying as a power law:
\begin{equation}
  P_\alpha(v) \sim |v|^{-(1+\alpha)}
   \qquad \mbox{for}\;|v|\gg 1\,,
   \label{eq:power}
\end{equation}
If $0 < \alpha < 2$, $v(t)$'s are in the basin of attraction of the 
$\alpha$-L\'evy stable distribution.
Since $\langle v^2 \rangle =\infty$, one has 
$\langle{|x(t)|^2}\rangle = \infty$. 
However introducing smaller moments 
$\langle{|x(t)|^q}\rangle = C_q t^{q/\alpha}$ with $q<\alpha$,
a typical position $X_f=\sqrt[q]{\langle{|x(t)|^q}\rangle}$
can be defined, which grows as $t^{\nu}$ with $\nu={1/\alpha}>{1/2}$.
This process, called L\'evy flight, is, therefore, superdiffusive.

\begin{figure}[!h]
{\centerline{
\includegraphics[width=8cm]{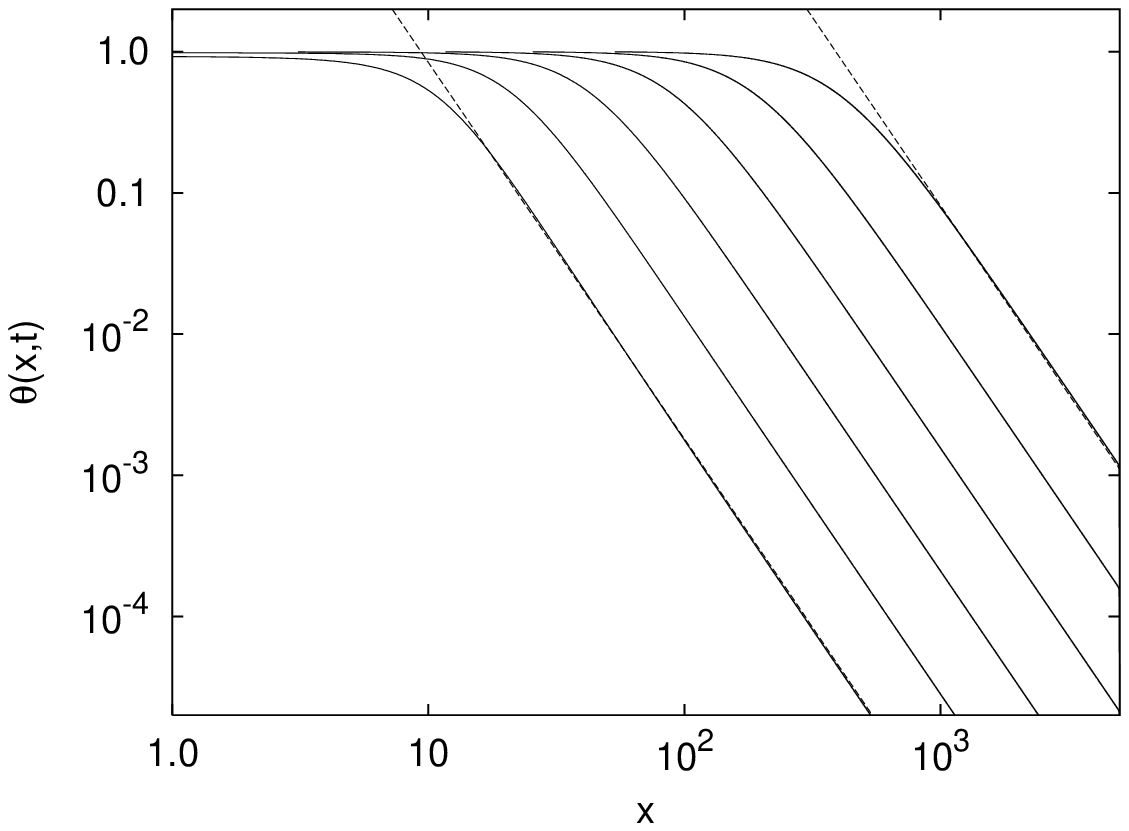}
\includegraphics[width=8cm]{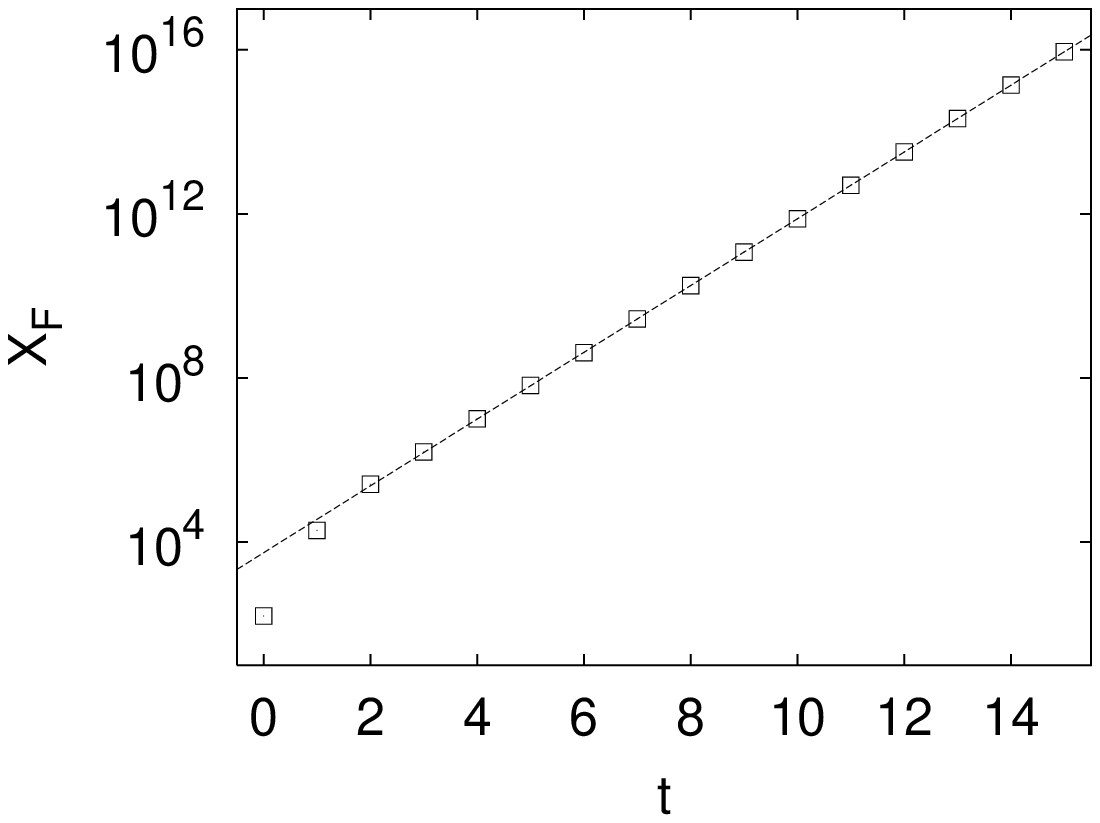}}}
\caption{Front shapes at increasing times in the case of $P(w) \sim
|w|^{-(1+\alpha)}$ with $\alpha ={5 \over 3}$ 
and $G(\theta)=\theta + {h \over \tau} g(\theta)$ with 
$\tau=1$ (left).
The dashed lines indicate the theoretical prediction $x^{-(1+\alpha)}$.
On the right it is shown the exponential growth of the typical
front position $X_f$ and its asymptotic behaviour proportional 
to $a^{t/(1+\alpha)}$, with $a=G'(0)$ (right). Adapted from~\cite{mancinelli}.
\label{fig:Levy_Fronts}
} 
\end{figure}

\begin{figure}[!h]
\centering\includegraphics[width=0.64\textwidth]{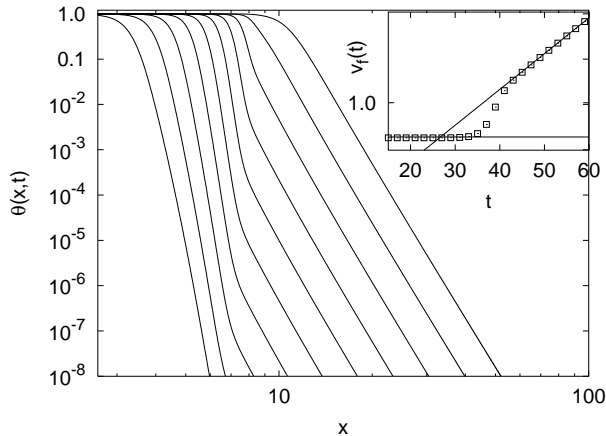}
\caption{Front shapes at $t=19, 22, 25 \cdots$ where $P(w)$ is the
basin of attraction of $P_\alpha(w)\sim |w|^{-(1+\alpha)}$
with $\alpha = 10$.
The inset shows the front speed $v_f(t) = X_f(t+1)-X_f(t)$. 
At short time, $v_f$ is constant as in the FKPP propagation,
whereas at longer time, the exponential propagation regime, 
$v_f \propto a^{\left[{t \over 1+\alpha}\right ]}$, takes place.
Adapted from~\cite{mancinelli}.
\label{fig:figgaus}
}
\end{figure}

Let us consider the reaction diffusion \eqref{eq:genFKPP}
where $\hat{L}$ corresponds to the process \eqref{eq:levy}. 
Only for simplicity in the notation and in the numerical computation
\cite{Abel}, we take into account a reacting term which is non zero
only at discrete time step, when $\delta$-form impulses occur:
\begin{equation}
   f(\theta,t) = \sum_{n=-\infty}^{\infty} g(\theta)
                                \delta(t - n h)\,h \,\,.
   \label{eq:reactmap}
\end{equation}
So between $0$ and $0^+$ Eq.~(\ref{eq:genFKPP}) can be easily 
integrated: $\theta(x, 0^+) = G(\theta(x,0))$
where $G(\theta)=\theta + {h \over \tau} g(\theta)$.
Then, between $0^+$ and $h$, Eq.~(\ref{eq:genFKPP}) reduces
to the linear equation $\partial_t \theta = \hat{L} \theta$,
and we can compute $\theta(x,h)$:
\begin{equation}
   \theta(x,h) = \int_{-\infty}^{+\infty} {\mathrm d}w
        P_{\alpha}(w) \theta(x - w, 0^+) 
        = \int_{-\infty}^{+\infty} {\mathrm d}w 
	P_{\alpha}(w) G(\theta(x-w,0))\,.
   \label{eq:integ}
\end{equation}
Iterating the previous procedure we can get $\theta(x,t)$, with $t = n
h$.  Let us note that Eq.~\eqref{eq:integ} is exact only when 
Eq.~\eqref{eq:reactmap} holds, while it is a good approximation when
$f(\theta)$ is not a $\delta$-impulsed function as long as $h$ is
small. In any case we are interested in pulled reactions (we recall
that ``pulled reactions'' means $G''(\theta) > 0$ and $G'(0) >
1$). For this family of reactions the detailed shape of $G(\theta)$ is
not important~\cite{Abel}.  Numerical results obtained integrating
Eq.~(\ref{eq:integ}), is shown in Fig.~\ref{fig:Levy_Fronts}.
\\
%-----------------------------FIG.7 -----------------------------------
%----------------------------------------------------------------------
Note that $X_f$ grows  exponentially fast in time, and
$\theta(x,t)$ develops a power law tail for $|x|\rightarrow \infty$.
The results shown in Fig.~\ref{fig:Levy_Fronts} can be
supported by an analytical argument (very similar to the one used for
diffusion process with exponential law tail PDF) via a linear
analysis, which is expected to be valid in the FKPP reaction case,
and the theory of the infinitely
divisible distributions.  For $\theta$ around zero, $G'(\theta)$ has a
linear shape, $G'(\theta) \simeq a \theta$ with $a>1$.  
Plugging this into Eq.~(\ref{eq:integ}) for $x\gg 1$ 
\begin{equation}
\theta(x,t) \simeq a \, (P_{\alpha} * \theta)(x, t-1) \simeq 
                   a^t \, (P_{\alpha} * P_{\alpha} * 
\cdots  \theta)(x,0)\,\,,
\label{eq:idd}
\end{equation}
where, for the sake of notation simplicity, we assumed $h=1$ (this
corresponds to a rescaling of the time) and the symbol ``*'' indicates
the convolution operator and $P_\alpha$ is the probability
distribution for a single jump, that is in the basin of attraction of
the $\alpha$-L\'evy stable distribution, giving for $x \gg 1$ and
large $t$: $$\theta(x,t) \sim |x|^{-(1+\alpha)} \, a^t\,\,,$$ which
predicts the power law tail behaviour of the front shape. Then, $X_f$ can
be determined using $\theta(X_f(t),t) \sim 1$, which gives
\begin{equation}
  X_f \simeq a^{\left [ {t \over 1+\alpha} \right ]}\,\,.
  \label{eq:novar}
\end{equation}
Now we wonder what happens if the distribution $P_\alpha$
has a power law tail but it belongs to the basin of attraction of a 
Gaussian PDF, i.e., if $\alpha \geq 2$.\\
At a first glance, since the diffusion is normal, one could expect 
the same features of the FKPP equation. Numerical results of
Fig.~\ref{fig:figgaus} show a rather different scenario.
For each localized initial condition, $\theta_0(x,t)$, 
already at the first step, the front has a shape not steep
enough to allow the usual FKPP propagation. 
In fact, because of the reaction, the tail of $\theta$ increases
exponentially in time. 
As a consequence of the Gaussian core of $P(x,t)$, we expect that the
bulk of $\theta$ behaves in the FKPP way, but, at large time, the
exponential growth of the tail has the dominant role, 
see~\cite{mancinelli} for a more detailed discussion.
Fig.~\ref{fig:figgaus} shows how the exponential form of 
the front, initially moving with a constant velocity, is overcome 
by a power law tail that grows more and more as time increases.
In other words, the initial FKPP behaviour (constant $v_f$ and 
$\theta(x,t)$ with exponential decay) is replaced by the
exponential increasing of the inert material
and $\theta(x,t)$ with a power law tail.

%-------------------------- FIG.8 --------------------------------------
%-----------------------------------------------------------------------

%%%%%%%%%%%%%%%%%%%%%%%%%%%%%%%%%%%%%%%%%%%%%%%%%%%%%%%%%%%%%%%%%%%%%
\section{Reaction diffusion on graphs: relevance of topology
\label{sec:reacgraph}}
%%%%%%%%%%%%%%%%%%%%%%%%%%%%%%%%%%%%%%%%%%%%%%%%%%%%%%%%%%%%%%%%%%%%%
In Sec.~\ref{sec:diffgraph} we discussed the diffusion processes 
on graphs. It is natural to introduce here a reaction term
in the diffusion equation on graphs \eqref{eq:graphdiffusion}:
\begin{equation}
\frac{d \theta_i}{dt} = 
\sum_{j} W_{ij} \theta_j+{1 \over \tau}f(\theta_i)
\label{eq:rdgraph}
\end{equation}
where the time has been rescaled to have a unique characteristic time:
$\tau$.  We recall that $W_{ij}= A_{ij} - k_i \delta_{ij}$ is the
Laplacian on the graph, and $k_i$ is degree of the node $i$.

A common way to study the spreading of the reaction on a graph is to
consider an initial concentration $\{\theta_i\}$ which is zero apart a 
small set of neighbours nodes, and look at the time course of the 
percentage of the reaction products 
$M(t)= \frac{1}{N}\sum_{i=1}^N \theta_i(t)$.
Of course, in the case of a regular lattice (e.g. square) with dimension 
$d=1, 2$ or $3$ one has
\begin{equation}
M(t)= { 1 \over N} \sum_{i=1}^N \theta_i(t)  \sim t^d \,\, ,
\label{eq:M_of_t}
\end{equation}
corresponding to a linear growth in a d-dimensional space of the 
initial condition. Let us wonder how the previous scaling is modified
in a reaction-diffusion process on graph.

%>>>>>>>>>>>>>>>>>>>>>>>>>>>>>>>>>>>>>>>>>>>>>>>>>>>>>>>>>>>>>>>>>>>>>>
\subsection{Reaction on geometrical graphs}
\begin{figure}
\centering
\includegraphics[scale=0.32]{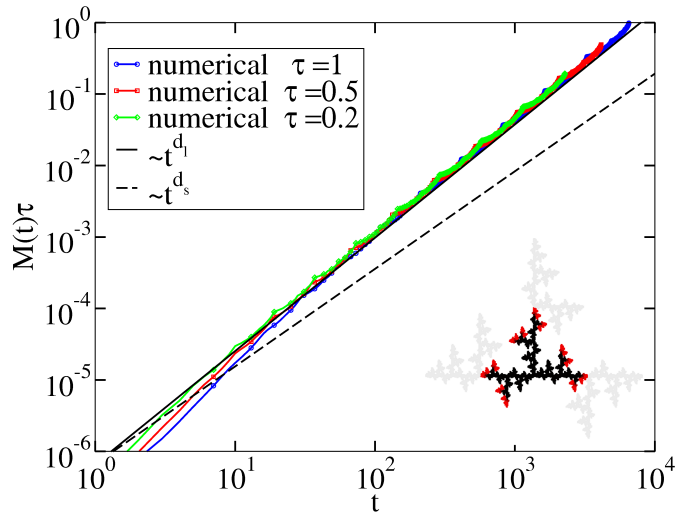}
\caption{The percentage of quantity of product times $\tau$, $M(t)\tau$
vs $t$. Numerical results for Eq.~\eqref{eq:rdgraph} 
are compared to prediction $t^{d_l}$ and $t^{d_s}$ in order to show
that, definitively, the spreading follows the connectivity dimension.
In the inset there is a graphical representation of the spreading
on a T-graph: the black part is burned material ($\theta=1$)
while the grey is the fresh one ($\theta=0$). Adapted from~\cite{BCVV}.}
\label{fig:graph}
\end{figure}

%>>>>>>>>>>>>>>>>>>>>>>>>>>>>>>>>>>>>>>>>>>>>>>>>>>>>>>>>>>>>>>>>>>>>>>
In a geometric graph, i.e. a graph in which each vertex is a point in
a space equipped with a metric, we expect that Eq.~\eqref{eq:M_of_t}
is still valid but with a different $d$ that is not so obvious to
identify.  In Sec.~\ref{sec:diffgraph} we introduced the fractal
dimension $d_f$ and the spectral dimension $d_s$, two quantities that
determine the scaling for the (typically anomalous) diffusion $\langle
x^2(t)\rangle \sim t^{d_s/d_f}$ with $d_s/d_f \le 1$ (see
Eq.~\eqref{eq:msd_graph}).  Detailed numerical simulations~\cite{BCVV}
on several graphs show
\begin{equation}
M(t) \sim t^{d_l}
\label{eq:M_tscale}
\end{equation}
where $d_l$ is the connectivity dimension (also known as 
``chemical dimension'') that measures the average number
of vertices connected to a vertex in at most $l$ link, as
$\#(l) \sim l^{d_l}$. As an example in Fig.~\ref{fig:graph} 
it is shown the scaling~\eqref{eq:M_tscale} for the T-graph 
for different reaction times $\tau$. A T-graph is a fractal
generated by iterating a rule which replaces a segment with a 
T-structure (from ``|'' to ``$\top$'', see Fig. 1 of 
ref.~\cite{Agliari}).
%------------------------ FIG.9 ----------------------------------
%-----------------------------------------------------------------
An easy way to validate such a behaviour is the following.  Denote
with $S_n$ the number of distinct sites visited by $n$ independent
random walkers, starting from the site $0$, after $t$ steps
$$
S_n(t)=\sum_{j=0}^N(1-C_{0j}(t)^n)
$$
where $C_{0j}(t)$ is the probability
that a walker starting from site $0$ has not visited site $j$ at
time $t$.
When the number of walkers is large ($n \to \infty$), $ C_{0j}(t)^n$ tends 
to zero if site $j$ has a nonzero probability of being reached in $t$ steps. 
In this limit, $S_n(t)$ represents all the sites which have nonzero
probability of being visited in $t$ steps and, $S_n(t) \sim t^{d_l}$. 
This is precisely the regime observed in the reaction spreading.

Let us note that  the result \eqref{eq:M_tscale} is another way to understand 
the standard front propagation in the comb lattice discussed in 
Sect.~\ref{sec:diffgraph}.
It is easy to realize that for such a graph $d_l=2$, therefore, in spite of 
the non homogeneous spatial structure, the spreading of the reactive material 
has the same scaling law than that of a simple regular $2D$ lattice.

%>>>>>>>>>>>>>>>>>>>>>>>>>>>>>>>>>>>>>>>>>>>>>>>>>>>>>>>>>>>>>>>>>>>>>
\subsection{Reaction on Erd\"os-Renyi graphs}
\begin{figure}
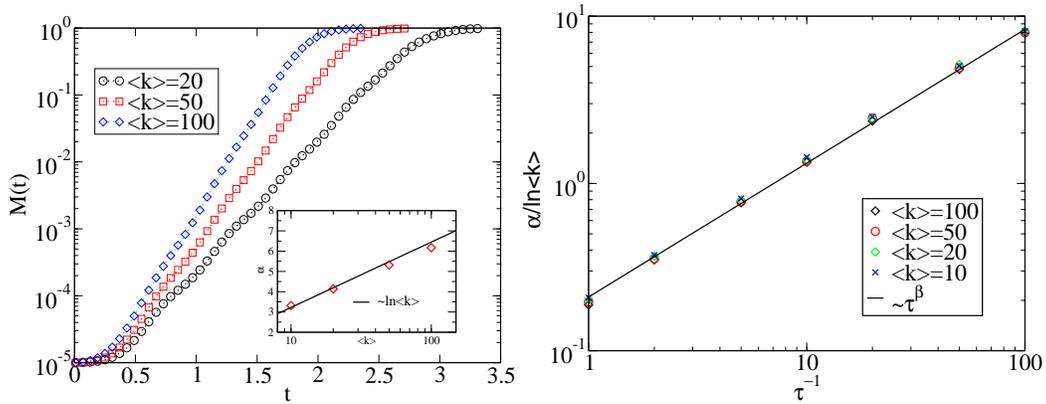

{\centering
\includegraphics[scale=0.25]{Fig_10a.eps}
\includegraphics[scale=0.25]{Fig_10b.eps}}
\caption{Reaction process Erd\"{o}s-Renyi graphs.
On the left it is reported $M(t)$ vs $t$ for three different average
degree of connectivity.  The reaction spreading follows an exponential
behaviour $M(t)\sim e^{\alpha t}$, where $\alpha$ depends on
$\langle{k}\rangle$ as shown in the inset.  On the right the scaling
exponent $\alpha$ normalized with $\ln\langle{k}\rangle$, as a
function of the inverse of reaction-time $\frac{1}{\tau}$, is
reported.  The straight line indicate $\tau^\beta$ with $\beta=-0.8$.
Adapted from~\cite{BCVV}.
}
\label{fig:ERgraph}
\end{figure}

%>>>>>>>>>>>>>>>>>>>>>>>>>>>>>>>>>>>>>>>>>>>>>>>>>>>>>>>>>>>>>>>>>>>>>
Let us now briefly discuss the spreading of the reaction on an
important class of graphs: the Erd\"os-Renyi (ER)
graphs~\cite{bollobas1998} characterized by $d_l=\infty$.  In the ER
graphs with $N$ vertices, two vertices are connected with probability
$p$ and the average degree of the graph connectivity is 
$\langle k \rangle=p(N-1)$. 
If $ p> \ln(N)/N$ the graph contains a global connected component
\cite{bollobas1998}.
On ER graphs the number of points in a sphere of radius $l$ grows
exponentially, $N(l ) \sim e^{c l}$ , hence we expect a similar
behaviour for the spreading process: $$ M(t)\sim e^{\alpha t} $$ as
shown in the left panel of Fig.~\ref{fig:ERgraph}
%----------------------------- FIG.10 --------------------------------
%--------------------------------------------------------------

If $\langle k \rangle$ is large and the reaction is slow 
enough we observe a two-step mechanism: first there is a rapid diffusion
on the whole graph, then the reaction induces an increase
of $\theta_i$. This leads to a simple mean-field reaction dynamics,
$d \rho /dt=\rho(1-\rho)/\tau$,
where $\rho$  is the average value of  $\theta_i$ on the graph.
In this case $\alpha=1/\tau$ as clearly observed in
numerical simulations (not shown here).
In the much more interesting case of a fast reaction, at each
time step the number of sites invaded is proportional to the
average degree of the graph, so that after $t$ steps we have
$$
M(t) \sim (C_1\langle k\rangle)^t= e^{C_2 \ln\langle k\rangle t}
$$
leading to $\alpha \sim \ln \langle k \rangle$, see the inset 
in the left panel of Fig.~\ref{fig:ERgraph}.
Furthermore, at variance with the case of graphs with finite $d_l$,
in the case of fast reaction and FKPP reaction term,
$\tau$ plays an important role since  $C_2$ is a function of $\tau$, 
see \cite{BCVV}:
$$
\alpha \simeq C \tau^{\beta} \ln  \langle k\rangle
$$
with $\beta\simeq -0.8$, as showed in the right panel
of Figure \ref{fig:ERgraph}.

%%%%%%%%%%%%%%%%%%%%%%%%%%%%%%%%%%%%%%%%%%%%%%%%%%%%%%%%%%%%%%%%%%%%%
\section{Conclusions}
\label{sec:concl}
%%%%%%%%%%%%%%%%%%%%%%%%%%%%%%%%%%%%%%%%%%%%%%%%%%%%%%%%%%%%%%%%%%%%%
Reaction-diffusion equations describe systems where the interplay
between transport and reactions processes produces spatio-temporal
variations on species concentrations.

The simplest situation occurs when the transport process follows 
the Fick's law and an autocatalytic term accounts for the reaction
process. This is the reference case, also called FKPP dynamics, 
which admits travelling wave solutions with a constant speed.

However, in certain contexts, such as complex fluids, crowded
environments or constrained geometries, the transport process
cannot be described by a Fick's law, and the concentration 
of the reaction-free problem can undergo anomalous diffusion, 
i.e. $\langle |\mathbf x(t) - \mathbf x(0)|^2
\rangle \sim t^{2 \nu}$ with $\nu \neq 1/2$.

Accordingly, the PDF is no more Gaussian and one can conjecture that a
simple scaling hypothesis:
\begin{equation}
P(x,t) \sim t^{-\nu} F\Big(\frac{|x|}{t^{\nu}}
\Big)
\label{eq:Pconc}
\end{equation}
can be still verified.  However the above scenario, called {\it weak}
anomalous diffusion, is not always valid.  Indeed, there exist more
complex situations, the so called {\it strong} anomalous diffusion,
that cannot be characterized by a single exponent $\nu$ 
because the moments present multiscaling: 
$\langle |\mathbf x(t) - \mathbf x(0)|^q \rangle \sim t^{q \nu(q)}$, 
where $\nu(q)$ is a non constant function, and a full
spectrum of exponents is required.\\

In this paper first we analyzed few basic mechanisms leading to
anomalous diffusion (both in the weak or strong version). In
particular, we focused on those cases in which the anomaly can be a
consequence of: i) the peculiar properties of advecting fields; or
ii) specific characteristic of stochastic processes; or finally
iii) the confinement of the random walks on
graph-like structures. Then we have seen how anomalous behaviour
affects the reaction-diffusion process and changes the front
propagation dynamics with respect to the reference case of the FKPP
equation.

A linear analysis on the reaction-transport equation shows that,
remarkably, if the PDF of the concentration of the reaction free
problem satisfies the Flory scaling, that is $F(z)=\exp(- c
|z|^{\alpha})$ with $\alpha=1/(1-\nu)$, the FKPP scenario still
survives. Whereas, if the previous scaling is violated the front
behaviour depends on the shape of the tails of the PDF, and not only on
the value $\nu$ (even if $\nu=1/2$).  Remarkably we have the
counterintuitive result that the front propagation could be faster than
linear also in the case in which $\langle |\mathbf x(t) - \mathbf x(0)|^2
\rangle \sim t$.
\\
Finally we considered the spreading of reactants on a graph.  Although
the diffusion on graphs is ruled by the ratio between the fractal
and the spectral dimension, the growth of the percentage of the
reaction products on the graph is determined by a unique exponent, 
i.e. the connectivity dimension $d_l$, $M(t)\sim t^{d_l}$.
In the case of random (Erd\"os-Renyi) graphs, where formally
$d_l=\infty$, we showed an exponential growth characterized by the
average degree of the graph connectivity.

%%%%%%%%%%%%%%%%%%%%%%%%%%%%%%%%%%%%%%%%%%%%%%%%%%%%%%%%%%%%%%%%%%%%

\end{document}